\documentclass[10pt,letterpaper]{article}
\usepackage{opex3}
\usepackage{amsmath,amssymb,graphicx,subfigure}
\usepackage{cite}
\graphicspath{{EPS/}}
\usepackage[utf8]{inputenc}
\usepackage{srcltx}

\begin{document}

%
%

\title{Laser Doppler holographic microscopy in transmission: application to fish embryo imaging}

\author{Nicolas Verrier$^1$, Daniel Alexandre$^{1,2}$ and Michel Gross$^{1,*}$}

\address{$^1$Laboratoire Charles Coulomb - UMR 5221 CNRS-UM2 CC 026 Universit\'{e} Montpellier II Place Eug\`{e}ne
Bataillon
34095 Montpellier cedex, France\\
$^2$Montpellier Rio Imaging, Centre de Recherche en Biochimie Macromol\'{e}culaire - UMR 5237 CNRS-UM1-UM2 1919 route de Mende
 34293 Montpellier  Cedex 5,  France }

\email{michel.gross@univ-montp2.fr} 



\begin{abstract}
We have extended Laser Doppler holographic microscopy to transmission geometry.  The  technique is validated with  living fish embryos imaged by a modified upright bio-microcope.  By varying the frequency of  the holographic reference beam, and    the combination of frames used  to calculate the hologram,  multimodal imaging has been performed. Doppler images of the blood vessels for different Doppler shifts,  images where the flow direction is coded in RGB colors or   movies showing blood cells individual motion have been obtained as well. The ability to select the Fourier space zone that is used to calculate the signal,  makes the method quantitative.
\end{abstract}

\ocis{
(090.1995)  Digital holography;
(090.2880)   Holographic interferometry;
(170.3340)   Laser Doppler velocimetry;
(180.3170)   Interference microscopy;
(290.5850)   Scattering, particles;
(300.6310)   Spectroscopy, heterodyne
}

\bibliographystyle{osajnl}


\section{Introduction}

Blood flow imaging techniques are widely used in biomedical studies, since they can assess physiological processes or can be used for   early detection of disease ~\cite{Friedman1995,Pase2012,kur2012cellular}. However, many blood flow studies require, for imaging purposes, the use of a contrast agent, making the blood flow characterization invasive~\cite{Sakurada1978,Kanno1987}. Scanning Doppler imaging techniques can be considered to alleviate this issue, but due to the scanning step, acquisition of an image is a time consuming process~\cite{Yeh1964}. 
An overview of the main techniques has been proposed by J. D. Briers~\cite{Briers2001}. Among these,
one can outline two ways of  monitoring the  blood flow: performing measurements either in spatial domain (speckle analysis) or in temporal domain (Doppler analysis).
Blood flow monitoring has been demonstrated by Laser Speckle Contrast Analysis/Imaging (LASCA/LSCI)~\cite{Briers1996}. Here, spatial statistics of the dynamic speckle are used to obtain blood flow images~\cite{Dunn2012,Briers2013}. Improvement of the acquired constrast image have been achieved through exposure time optimization~\cite{YuanDevor2005}, or intensity fluctuation analysis~\cite{ZengWang2013} resulting in high quality perfusion images. However, these techniques are only limited to perfusion monitoring and do not allow to assess other quantities such as amplitude or phase contrasts.

Originally proposed by Gabor~\cite{Gabor1948} as an optimization of electron microscopy, classical holography aims at recording, on a photo-sensitive material, the interference between a reference field and a diffracted object field. However, within the proposed common-path interferometric configuration, holographic imaging suffer from the so-called twin-image noise~\cite{Gabor1949}. In 1962, Leith and Upatnieks propose to introduce an off-axis reference field, which lead to a separation of the object and its twin image contribution in spatial frequencies domain~\cite{LeithUpatnieks1962}.
Development and democratization of high resolution CCD/CMOS sensors paid a major role in holographic imaging spreading out, making possible both digital recording and processing of holographic images~\cite{Schnars1994}. Therefore, due to its intrinsic properties, digital holographic imaging is used in a wide variety of studies in fluid mechanics~\cite{Lozano1999,PuMeng2005,DessePicart2008,Verrier2008,Verrier2010}, biomedical imaging and microscopy ~\cite{Charriere2006,XuJericho2001,Kemper2008,kim2010principles,lee2013quantitative}, or mechanical inspection~\cite{PicartLeval2005,LevalPicart2005,Asundi2006}.

Further development led to the introduction of heterodyne digital holography~\cite{LeClerc2000}. Here, the reference field is dynamically phase shifted with respect to the object field. Therefore, the recorded hologram is time modulated, thus enabling phase-shifted interferometric measurements~\cite{AtlanGrossAbsil2007}. Moreover, this data acquisition scheme as been demonstrated to be shot-noise limited~\cite{VerpillatJoud2010,Lesaffre2013}. Temporal modulation feature of digital heterodyne holography can also be used to investigate dynamic phenomena, and being considered as a laser Doppler imaging technique~\cite{AtlanGrossRSI2007}. The ability of heterodyne digital holography to perform Doppler imaging has been demonstrated in various domains such as microfluidics~\cite{AtlanGrossLeng2006}, vibration motion characterization~\cite{VerpillatJoud2012,VerrierAtlan2013,VerrierGrossAtlan2013},  \emph{in vivo} vasculature assessment, without contrast agent~\cite{AtlanGrossForget2006,AtlanForgetBoccara2007,Simonutti2010} and motion of  biological objects \cite{fang2007imaging,Iwai2014_159,joseph_2010}.
%

In this paper, we combine laser Doppler holography ~\cite{AtlanGrossRSI2007} and transmission microscopy to analyze blood flow in fish embryos.
We have adapted a  laser Doppler holographic setup to a standard  bio-microscope by carrying the two beams of the holographic interferometer (illumination of the object and reference), whose  frequency offset is controlled, by optical fibers.
%
%
Multimodal acquisition and analysis of the data is made by acting on  the frequency offset of the two beams, and on the location of the Fourier space filtered zone.  With the same set of data,
we have imaged,  with amplitude contrast,  the whole fish embryo, or only  the moving blood vessels. In that last case, we have obtained images where the flow direction is coded in RGB color. For large vessels, the dependance of the holographic signal with the frequency offset yields the Doppler frequency shift, and thus the flow velocity. For small vessels, individual Red Blood Cells (RBC) can be imaged, and movies showing the RBC motion are obtained.

\section{Materials}

\subsection{Heterodyne holographic microscopy  setup}
\begin{figure}[htbp]
\centering
\includegraphics[width = 12 cm]{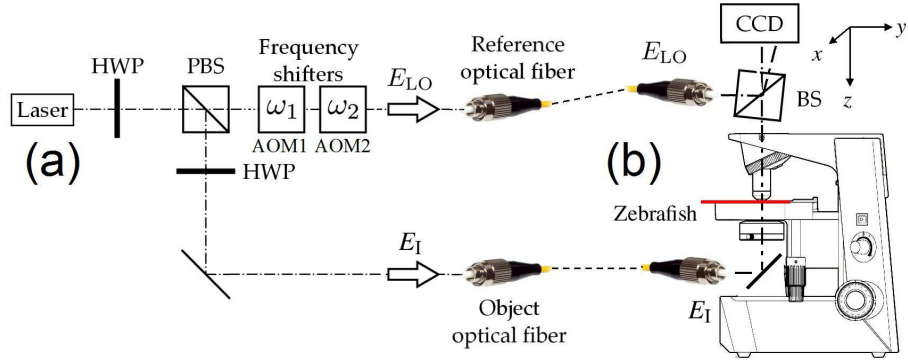}
\caption{Heterodyne digital holographic microscopy experimental arrangement (a) Injection part of the interferometer.  HWP: half wave plate; PBS: polarizing beam splitter; AOM1, AOM2: acousto optic modulators (Bragg cells). (b) Classical upright microscope used for the off-axis recombinaison of reference and object beams. BS: cube beam splitter; CCD: CCD camera.}\label{Setup}
\end{figure}
%

Our off-axis digital holography set-up is  a Mach-Zehnder interferometer in which the recombining cube beam splitter is angularly tilted. To  fit our set-up   into regular upright microscope,  the interferometer is split into an injection (a) and a recombination (b) part, which are connected together with optical fibers (Fig. \ref{Setup}).

The injection part  (Fig.  \ref{Setup}(a)) is mounted on a separate breadboard, and is  used to build both object illumination (field $E_{\rm I}$) and reference (or local oscillator (LO) $E_{\rm LO}$) beams. Laser light is provided by a 785 nm, 80 mW Sanyo\textregistered \ DL7140-201S continuous mono-mode laser diode (angular frequency $\omega_{\rm I}$). Laser beam is split by a polarizing beam splitter (PBS), which in combination with the first half wave plate (HWP) allows to adjust the optical power in both reference and object arms of the interferometer. To enable phase shifting interferometry and frequency scanning of the holographic detection frequency, we control the frequency  $\omega_{LO}$  of the reference beam by using the  heterodyne holography method \cite{LeClerc2000}.  The reference arm  is thus  frequency shifted by using two acousto-optic modulators AOM1 and AOM2 (80 MHz AA electronics\textregistered), operating  at $\omega_1$ and $\omega_2$ respectively,  so that $\omega_{LO}=\omega_{\rm I} + \Delta \omega$ with $\Delta\omega=\omega_2-\omega_1$.
Reference and object fields are finally injected into two mono-mode optical fibers (Thorlabs\textregistered \ P1-780A-FC-2 fiber patch).

Both reference ($E_{\rm LO}$) and object illumination ($E_{\rm I}$) fibers are brought to an upright microscope (Olympus\textregistered \ CX41) working in transmission configuration (Fig.  \ref{Setup}(b)). The object is imaged by an Olympus DPlan microscope objective (MO: NA= 0.25, G=10, corrected at 160 mm).  The illumination field ($E_{\rm I}$) is scattered by the studied living object (the zebrafish embryo) yielding the object signal field ($E$).  Signal ($E$) and reference field ($E_{LO}$) are combined together by a cube beam splitter (BS) that is angularly tilted thus introducing an off-axis angle in the interferometric arrangement. Interferences (i.e. $E+ e^{j\Delta \omega t} E_{LO}$) between both fields are recorded on a 1360$\times$1024 pixel (6.45 $\mu$m square pitch) 12-bits CCD camera (Jenoptik ProgRes MF\textregistered) operating at a framerate of $\omega_{\rm S} / (2\pi) \leq 10\ {\rm Hz}$. Recorded data are cropped to $1024\times 1024$ and FFT (Fast Fourier Transform) are made on a  $1024\times 1024$ calculation grid.
The CCD to MO distance is such that the effective MO enlargement factor is $G'=10.4$ (instead of $G=10$).
In order to perfectly control the phase and frequency of the $E_{LO}$ reference beam, the CCD camera is triggered at frequency $\omega_{CCD}$, and the $\omega_1$, $\omega_2$ and $\omega_{CCD}$  signals are generated by digital synthesizers driven by a common 10 MHz clock.

%

Zebrafish (Danio rerio, wild type AB line) were maintained according to standard protocols \cite{westerfield1995zebrafish}. Larvae were mounted on a standard glass slide under a 22 mm 1.5 round coverslip with a 0.5 mm thick caoutchouc ring  as a spacer and sealant. They were embeded in a drop of cooling 1.5\% low melting point agarose at 37$^\circ$C and oriented before gelation. The chamber was filled with 100 mg/l tricaine in filtered tank water and imaging was performed at room temperature.  Blood vessels nomenclature is according to \cite{isogai2001vascular}.

\section{Laser Doppler holography mode}

\begin{figure}[htbp]
\centering
\includegraphics[width = 8cm]{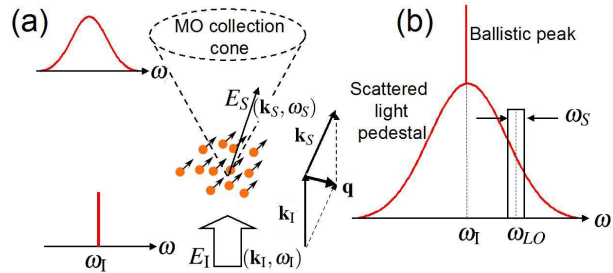}
\caption{Detection of zebrafish embryo blood flow in transmission geometry. (a) Illustration of the spectral broadening of light encountering moving scattering objects (here, red blood cells). (b) Principles of the spectral measurement by heterodyne digital holography.}\label{Broadening}
\end{figure}

Due to motions within the sample (mainly the RBC  motion  within the embryo vasculature),
the incident field $E_{\rm I}$ (of wave vector $\textbf{k}_{\rm I}$ ) that is  scattered
to wave vector $\textbf{k}_S$  (yielding $E_S$) undergoes a Doppler shift

\begin{eqnarray}\label{Eq_Doppler_shift}
  \omega_S-\omega_{\rm I}=\textbf{q} \textbf{ .} \textbf{v}
\end{eqnarray}
where $\textbf{v}$ is the velocity of the scatterer and $\textbf{\textbf{q}}=\textbf{k}_S -\textbf{k}_{\rm I} $ the scattering vector
(see Fig.  \ref{Broadening}(a)) and $\textbf{q} \textbf{.} \textbf{v}$ the inner product of vectors $\textbf{q}$ and $\textbf{v}$.
%
%
As detection is made for all wave vectors $\textbf{k}_S$ within the MO collection cone, the frequency spectrum of the moving scatterers  signal  is broadened. One then gets  two components: a sharp narrow peak from the  photons that are not scattered (ballistic photons) or scattered   by objects that do not move, and a broad pedestal from the photons that are scattered by moving objects and that are collected by MO. The  width of the pedestal is  proportional to the MO numerical aperture $\textrm{NA}$ since $|\textbf{q}| \sim  \textrm{NA} |\textbf{k}_{\rm I}|$. With $\textrm{NA}=0.25$, the Doppler broadening we observe here is much smaller  (10 to 100 Hz) than the one (100 to 1000 Hz) observed in previous \textit{in vivo} Doppler holographic  experiments  made in reflection configuration \cite{atlan2006frequency,atlan2007cortical,atlan2008high,Simonutti2010}, where $|\textbf{q}| \sim 2 |\textbf{k}_{\rm I}|$.
Since our camera operates at $\omega_{\rm S} / (2\pi) = 10\ {\rm Hz}$, the  bandwidth of our holographic detection ($\pm 5$ Hz around $\omega_{LO}/(2\pi)$) in not enough to acquire the moving scatterer Doppler signal.  As in previous works, the frequency  $\omega_{LO}=\omega_{\rm I}+\Delta\omega$  (with $\Delta\omega=\omega_{1}-\omega_{2}$) is swept  in order to acquire the whole Doppler signal and to analyze its  spectrum.

\section{Experimental results}

\begin{figure}[htbp]
\centering
\includegraphics[width = 4cm]{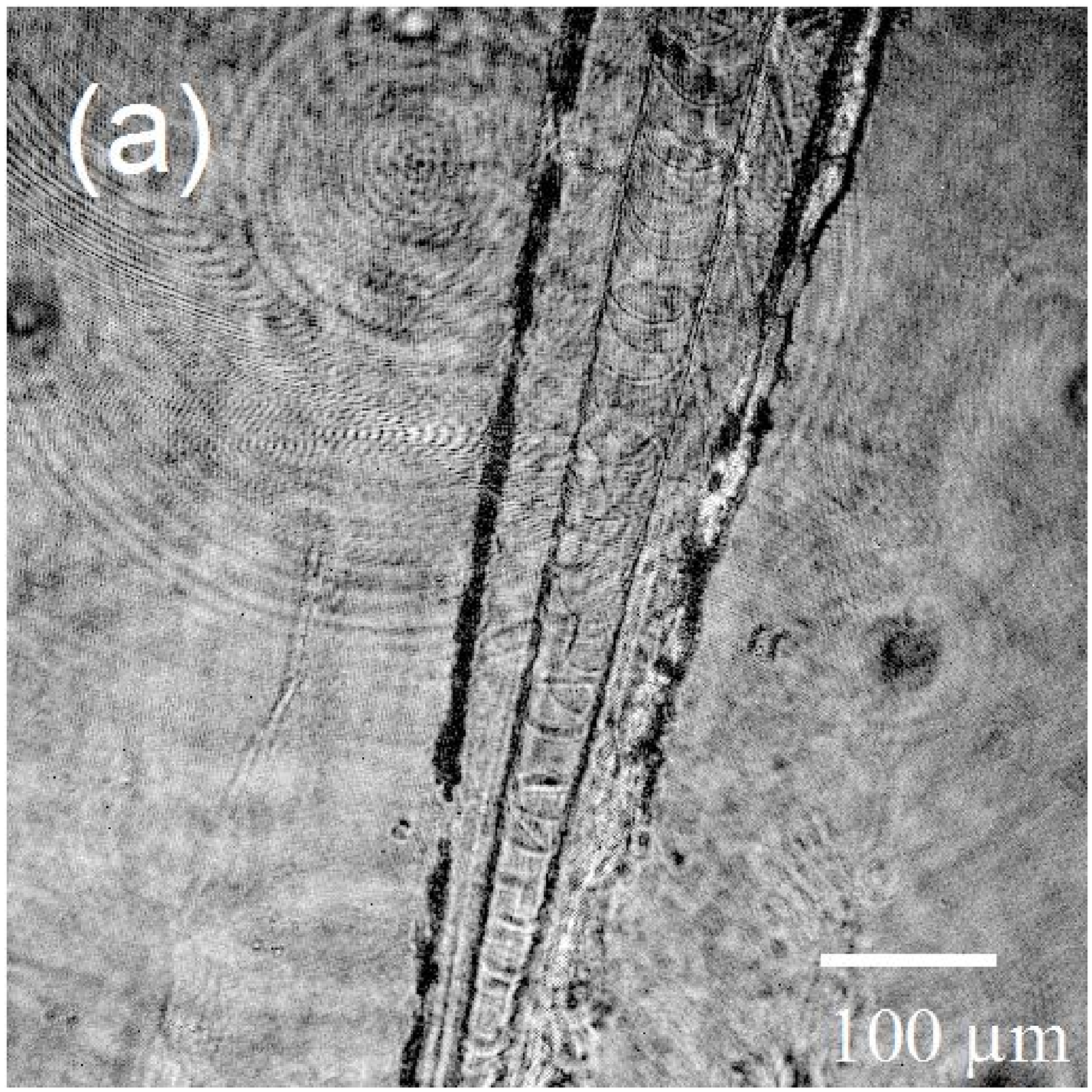}
\includegraphics[width = 4cm]{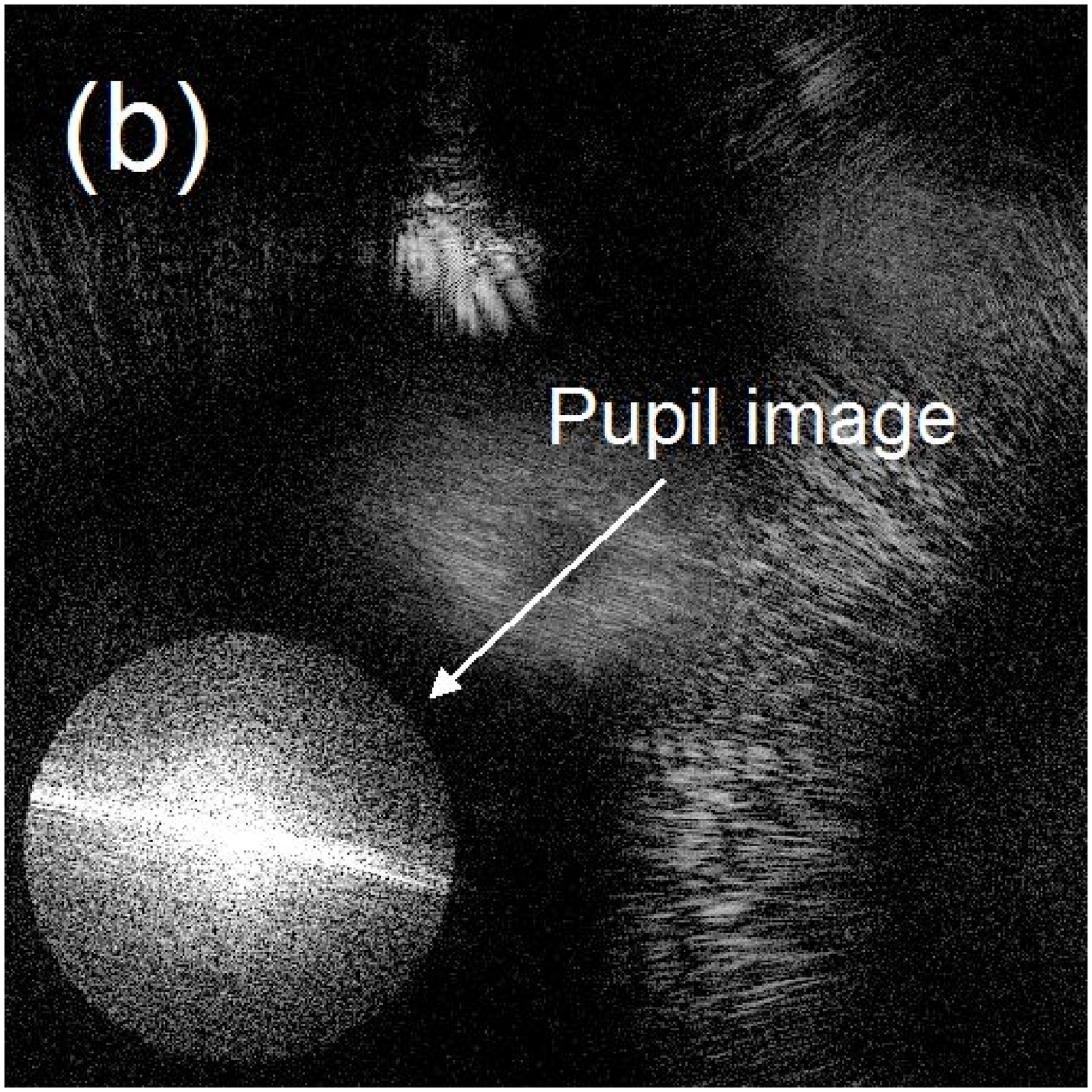}\\
\includegraphics[width = 4cm]{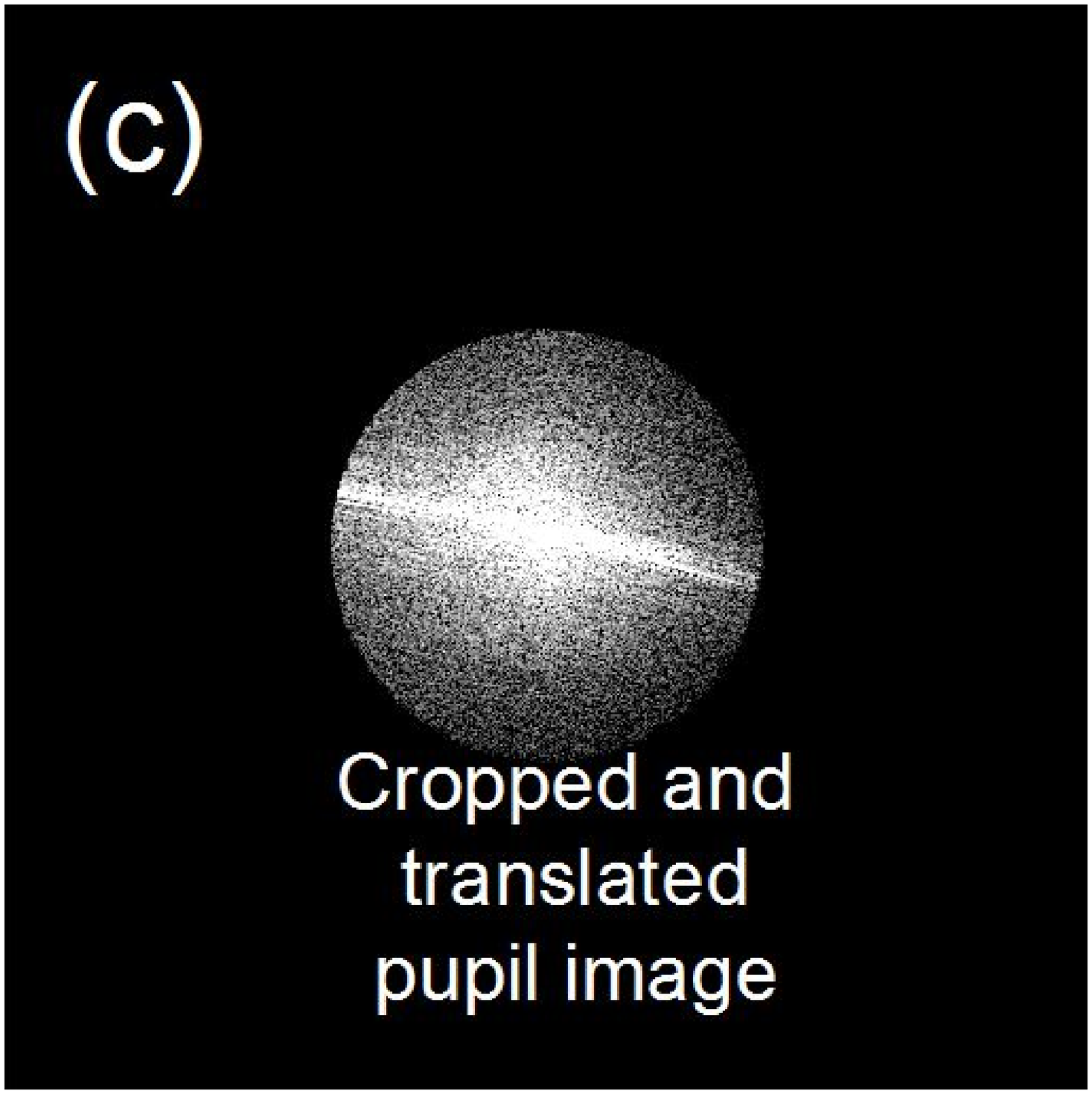}
\includegraphics[width = 4cm]{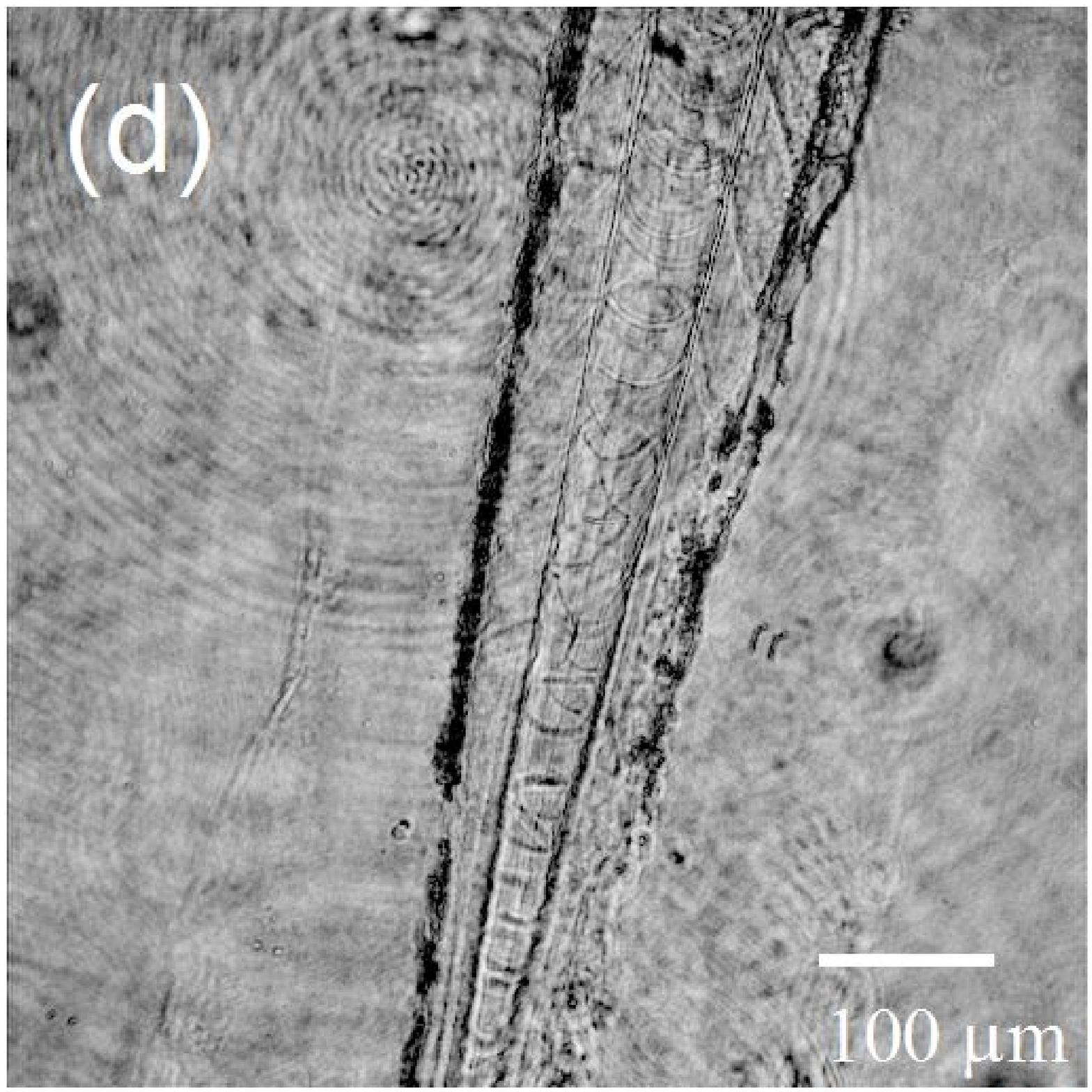}
\caption{Spatial filtering and reconstruction principle.  Holograms $H(x,y)$ (a), $H_1(k_x,k_y)$ (b), $H_2(k_x,k_y)$ (c) and $H_3(x,y,z=0)$ (d): see \url{Media 1}. The display is made in arbitrary Log scale for the average intensity  $\langle |H_{X}|^2\rangle$. This average is calculated  made by making the reconstruction with $I_0..I_3$, with $I_1..I_4$... and with  $I_{28}..I_{31}$, and by averaging the resulting $|H_{X}|^2$. Images (a) and (d) show  a side view of a 6 days old zebrafish embryo.  Ventral side in on the right and anterior is up. }\label{fig_BW}
\end{figure}

The  holographic information is acquired by sweeping the frequency shift  $\Delta \omega /(2\pi)$ from  $0$  up to $\sim 100$ Hz by step of $\omega_{CCD}/2\pi$ (i.e. $\Delta \omega=m \omega_{CCD}$ with $m$  integer) except in section \ref{Section_reconstrcution} where $\Delta \omega=\omega_{CCD}/4$. On the other hand, the movies \url{Media 1} and \url{Media 2} have been recorded with $\Delta \omega=0$.  To optimize sensitivity and selectivity,
the illumination beam versus local oscillator ratio is adjusted in order to have $|E_{LO}|^2 >|E|^2$,
with $|E|^2$ as large as possible. Exposure time and frame rate are   $T_{exp}=50$ ms and $\omega_{CCD}/(2\pi) = 1/T_{CCD} = 10$ Hz.
For each frequency shift $\Delta \omega /(2\pi)$, sequences of $N=32$ (or $N=96$) CCD camera frames (i.e. $I_0$, $I_1$...$I_{N-1}$)
are recorded.

\subsection{Spatial filtering and reconstruction.}\label{Section_reconstrcution}

The reconstruction procedure  is similar to the one used in previous works \cite{warnasooriya2010imaging,verpillat2011dark}. To illustrate reconstruction, let us consider 4 phase detection of the ballistic peak signal (no Doppler shift). We have thus adjusted $\omega_1$ and $\omega_2$ to have  $\Delta \omega=\omega_{CCD}/4$, and calculated the demodulated hologram by:
\begin{eqnarray}\label{Eq_H_4phases}
  H(x,y) &=& \left[I_0(x,y)-I_2(x,y)\right]~+~j\left[I_1(x,y)-I_3(x,y)\right]
\end{eqnarray}
Figure \ref{fig_BW}(a) shows the holographic signal $|H(x,y)|^2$ displayed in log scale. To select the +1 holographic grating order by the Cuche et al. method \cite{Cuche2000}, we have performed the holographic reconstruction  of the MO pupil  by the Schnars et al. method \cite{Schnars1994} yielding $H_1(k_x,k_y)$ with:
\begin{eqnarray}\label{Eq_H1}
  H_1(k_x,k_y) &=& \textrm{FFT}\left[~ H(x,y) ~e^{j |\textbf{k}| (x^2+y^2)/2d} ~\right]
\end{eqnarray}
where $\textrm{FFT}$ is the Fast Fourier transform operator, and $e^{j |\textbf{k}| (x^2+y^2)/2d}$ the phase factor that makes the pupil image sharp (see Fig. \ref{fig_BW}(b)). For a plane wave reference, the parameter $d$ is equal to the MO pupil to CCD camera distance. Optimal spatial filtering is then made by cropping the MO pupil disk on its sharp circular  edge, and by moving the cropped zone in the center of the Fourier space  (see Fig. \ref{fig_BW}(c)) yielding  $H_2(k_x,k_y)$. The reconstructed hologram  $H_3(x,y,z)$ is then calculated by the angular spectrum method that involves 2 FFT from $\textrm{FFT}^{-1}[H_2(k_x,k_y)]$. We have thus:
 \begin{eqnarray}\label{Eq_H3}
   H_3(x,y,z) &=& \textrm{FFT}^{-1}\left[~H_2(k_x,k_y)~e^{j(k_x^2+k_y^2)/2z}~\right]
 \end{eqnarray}
where ~$e^{j(k_x^2+k_y^2)/2z}$ is the phase factor that describe the propagation of the field from the CCD plane $z=0$ to the location $z$ of the zebrafish image (via MO). In most cases, the embryo is in focus on the camera, and $z=0$. Here, the selection of the +1 grating order is made by the crop, the compensation of the MO phase curvature  by the $e^{j |\textbf{k}| (x^2+y^2)/2d}$  phase factor, and the compensation of the off axis tilt by the translation of the cropped zone.


We have calculated the so called instantaneous intensity signal  $|H_{3}|^2$ by using data of successive sequences of 4 frames, i.e. by calculating $|H_{3}|^2$  with $I_0..I_3$, then with  $I_1..I_4$... and to the end with $I_{28}..I_{31}$). The movie of the instantaneous intensity signal $|H_{3}|^2$ (see \url{Media 1}) shows that  the time varying components of  $|H_{3}(t)|^2$ are extremely low.
We have then averaged $|H_{3}(t)|^2$ over the sequence of $n$ frames, getting    $\langle |H_{3}|^2\rangle$ that is displayed on Fig. \ref{fig_BW}(d).

The images show a side view of the zebrafish embryo.  Due to the use of coherent light illumination, the  image exhibits some speckle noise but anatomical details such as the notochord, somite boundaries and pigment cells are distinguished. In the corresponding movie \url{Media 1}, motion of blood can be barely seen in the caudal artery and vein  but none is observe in the smaller capillaries.

\subsection{Holographic Doppler images of the moving scatterers.}

To select the moving scatterer signal  that corresponds to the Doppler pedestal, we have swept the detection frequency shift $\Delta \omega$  by step $\omega_{CCD}$, and calculated  2 phase holograms from the recorded data. We have thus:
\begin{eqnarray}\label{Eq_2phase}
 \Delta \omega&=& \omega_{LO}-\omega_{\rm I} = m \omega_{CCD} \\
\nonumber  H(x,y) &=& I_0(x,y)-I_1(x,y)
\end{eqnarray}
where $m$ is integer. Note that  $ H(x,y)$ is defined by Eq. \ref{Eq_H_4phases} for 4 phase holograms, and  by Eq. \ref{Eq_2phase}  for 2 phase holograms. We have kept the same notation (i.e. $H(x,y)$), because the reconstruction equations Eq. \ref{Eq_H1} and  Eq. \ref{Eq_H3}  are  the same in both cases.

\begin{figure}[htbp]
\centering
\includegraphics[width = 7cm]{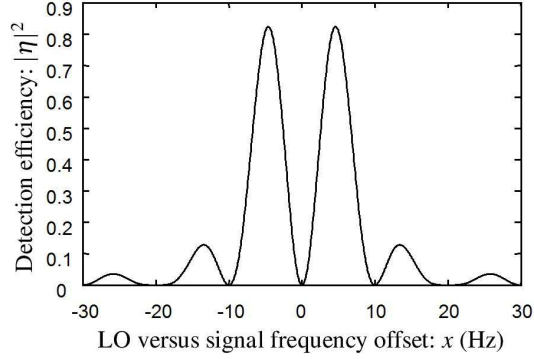}
\caption{Two phase detection efficiency $|\eta|^2$  as a function of the LO versus signal frequency offset: $x$. Calculation is made by Eq.\ref{Eq_ETA} with $T_{CCD}=100$ ms and $T_{exp}=50$ ms.}
\label{fig_eta2}
\end{figure}

We have calculated the 2 phase holographic detection efficiency (i.e. $\eta$ for the field and $|\eta|^2$ for the intensity) as a function of the frequency $\omega_S$ of  the scattered field. This calculation is similar  to the one done in previous works \cite{AtlanGrossAbsil2007,VerpillatJoud2010}.  We get:
\begin{eqnarray}\label{Eq_ETA}
     \eta(x) &=& \frac{1}{2 T_{exp}} \sum_{k=0}^1~(-1)^k \int_{t=k T_{CCD}-T_{exp}/2}^{k T_{CCD}+T_{exp}/2} e^{j 2 \pi x t}~dt\\
     \nonumber &=& \frac{1}{2} \textrm{sinc}(\pi x T_{exp}) \left[ 1-  e^{j 2\pi  x T_{CCD}}\right]
\end{eqnarray}
where $T_{CCD}=2\pi/\omega_{CCD}$ is the frame time interval, and   $x=(\omega_S-\omega_{LO})/2\pi$ the LO versus signal frequency offset. The efficiency $\eta^2$ is plotted on Fig.\ref{fig_eta2}. As seen, the 2 phase detection selects scattered field components with $x$ close to zero (i.e. with   $\omega_S \simeq \omega_{LO}$). Moreover, since $|\eta|^2$ is null for $x=-20,-10,0,10,20...$ Hz, the holographic detection efficiency $|\eta|^2$ remains null at the illumination frequency $\omega_{\rm I} $,  whatever the frequency shift $ \Delta \omega=m \omega_{CCD} $  is. The ballistic peak (whose frequency is $\omega_I$), which is much bigger than the Doppler pedestal, is thus not seen during the sweep.

\begin{figure}[htbp]
\centering
\includegraphics[width = 3.4cm]{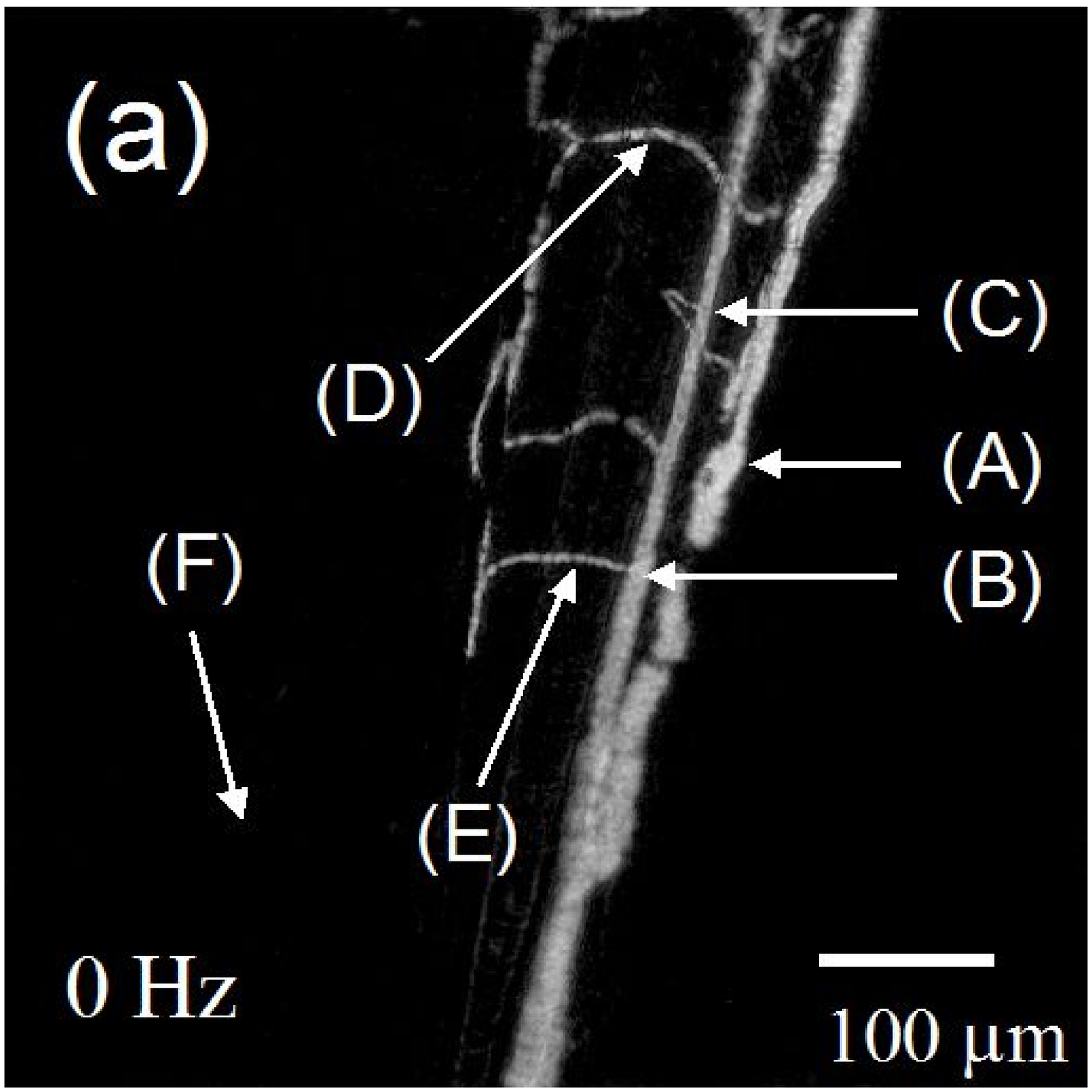}
\includegraphics[width = 3.4cm]{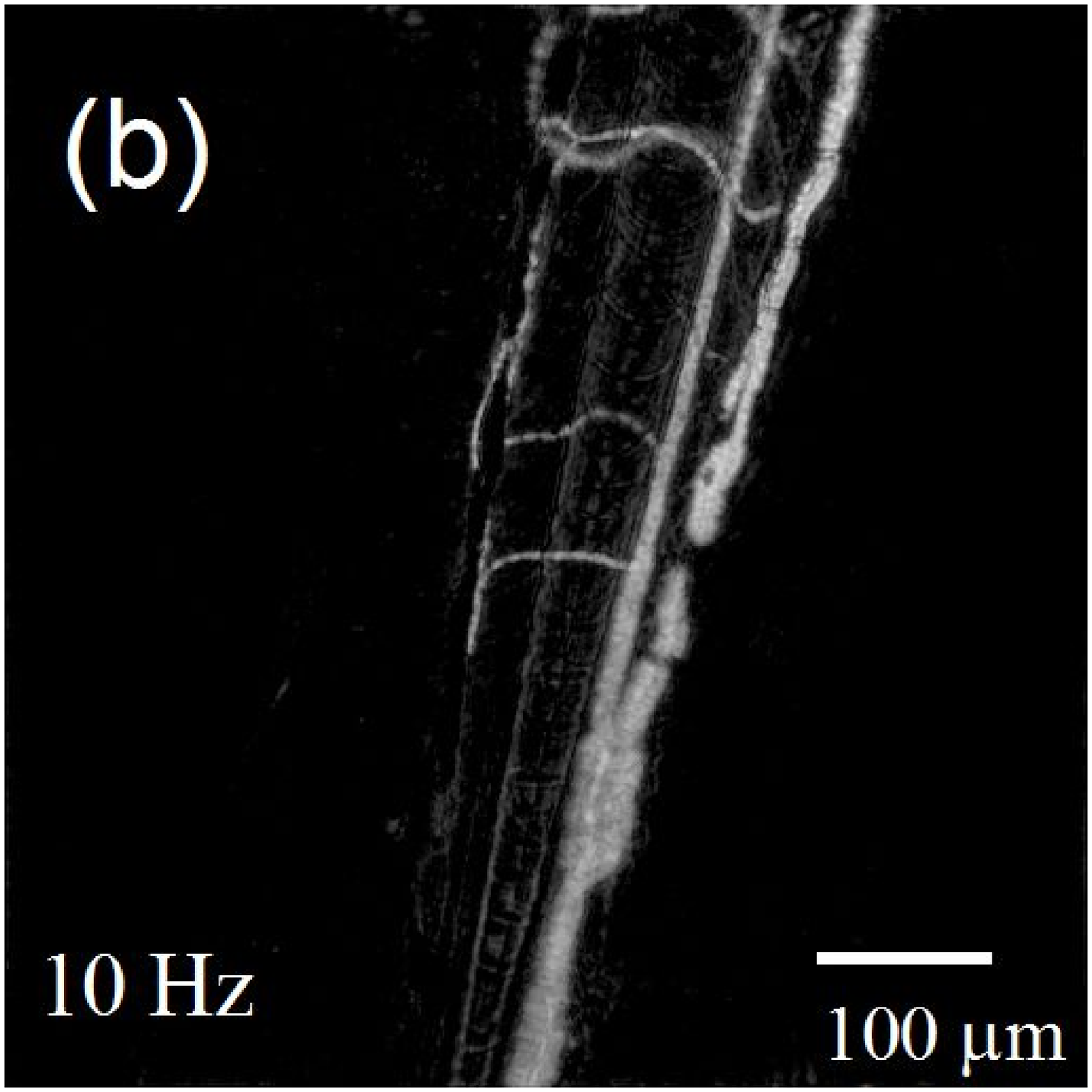}
\includegraphics[width = 3.4cm]{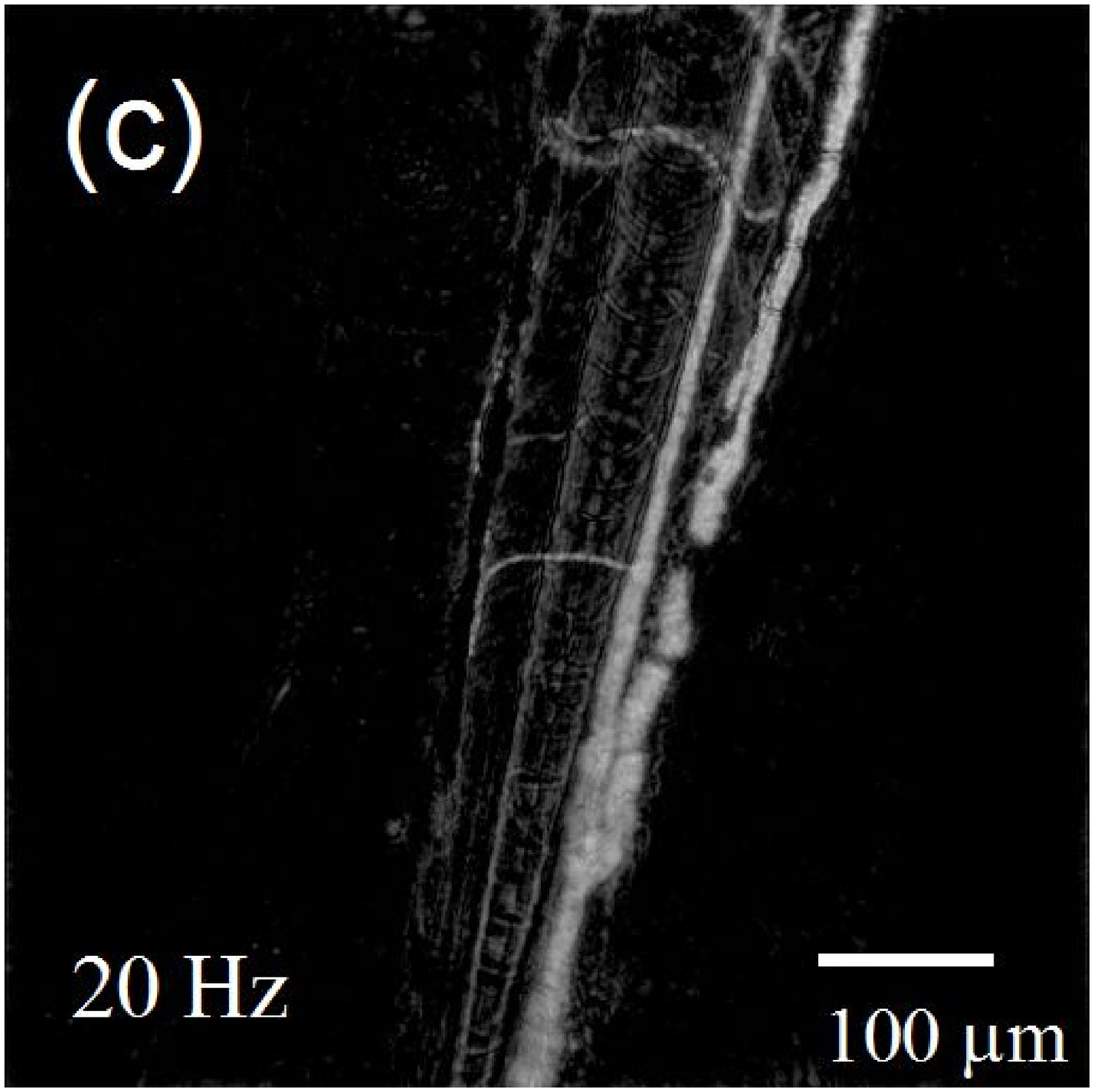}
\includegraphics[width = 3.4cm]{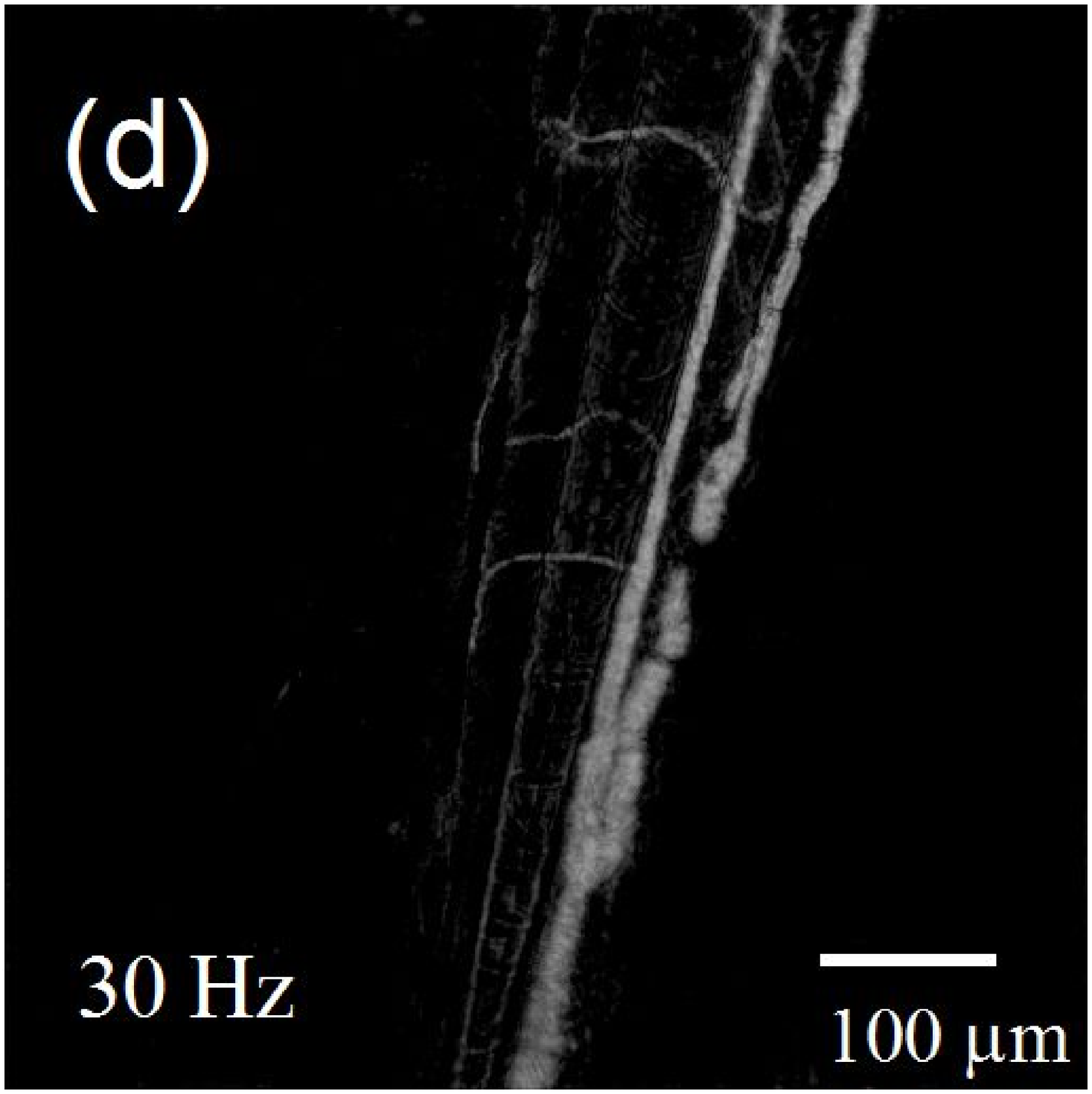}
\includegraphics[width = 3.4cm]{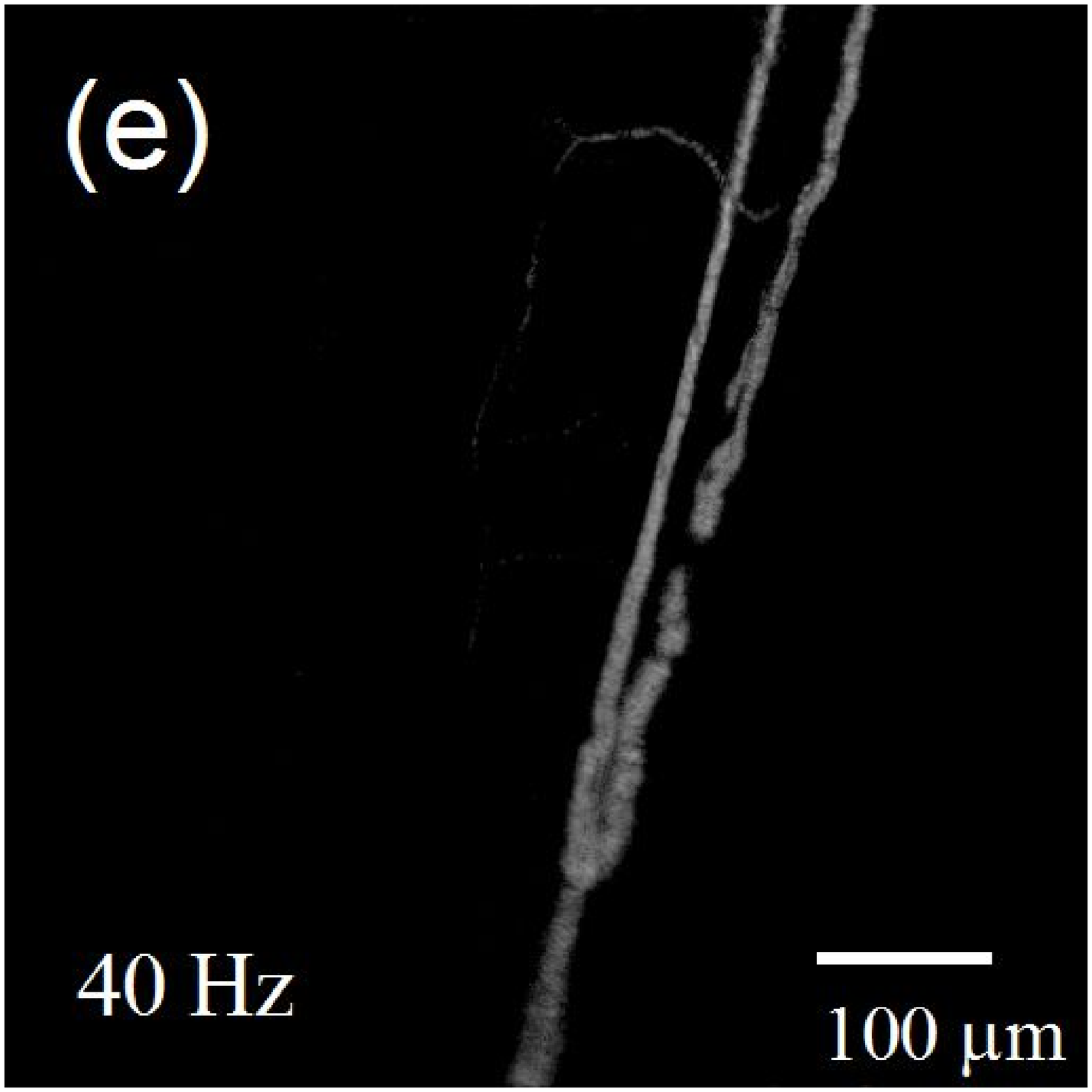}
\includegraphics[width = 3.4cm]{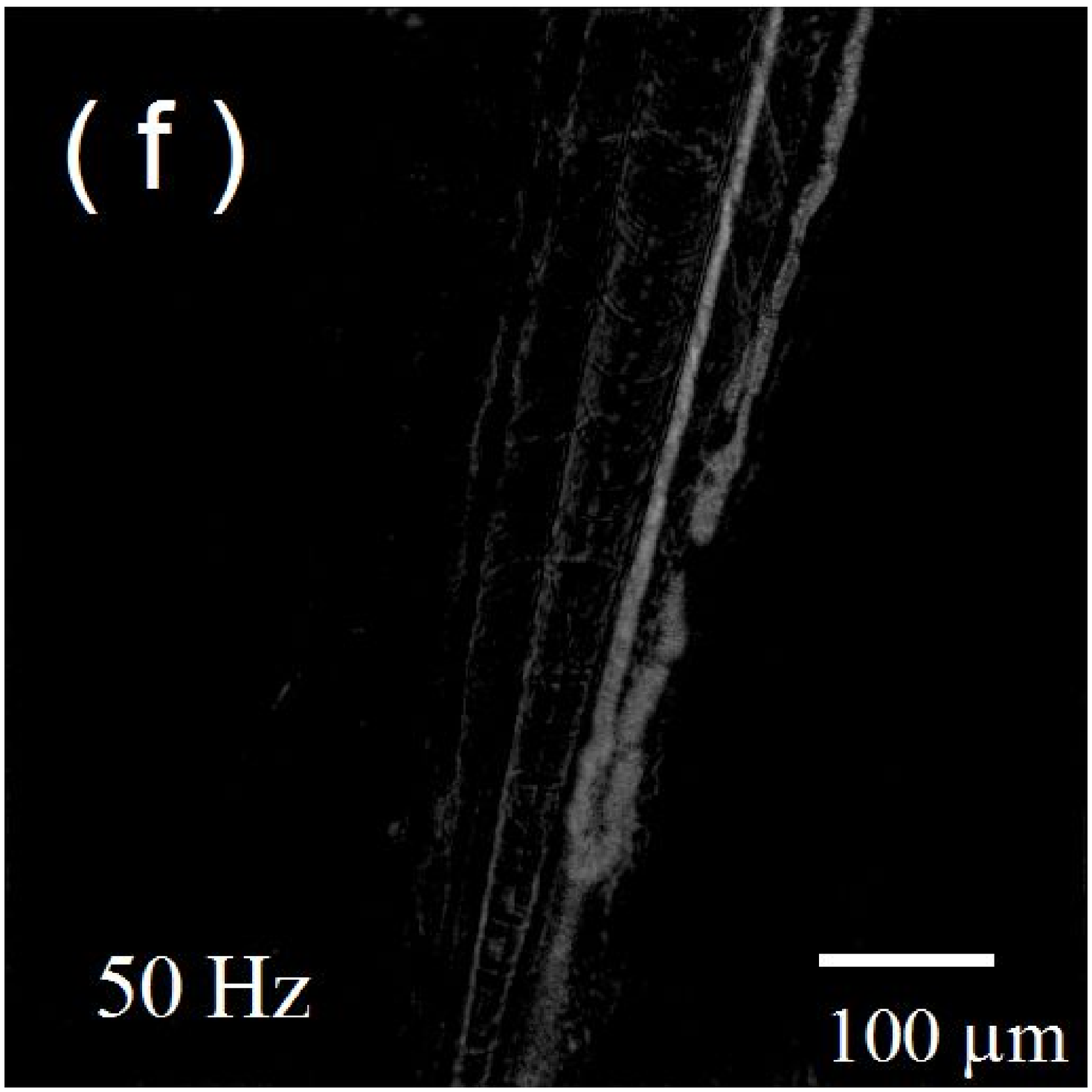}
\includegraphics[width = 3.4cm]{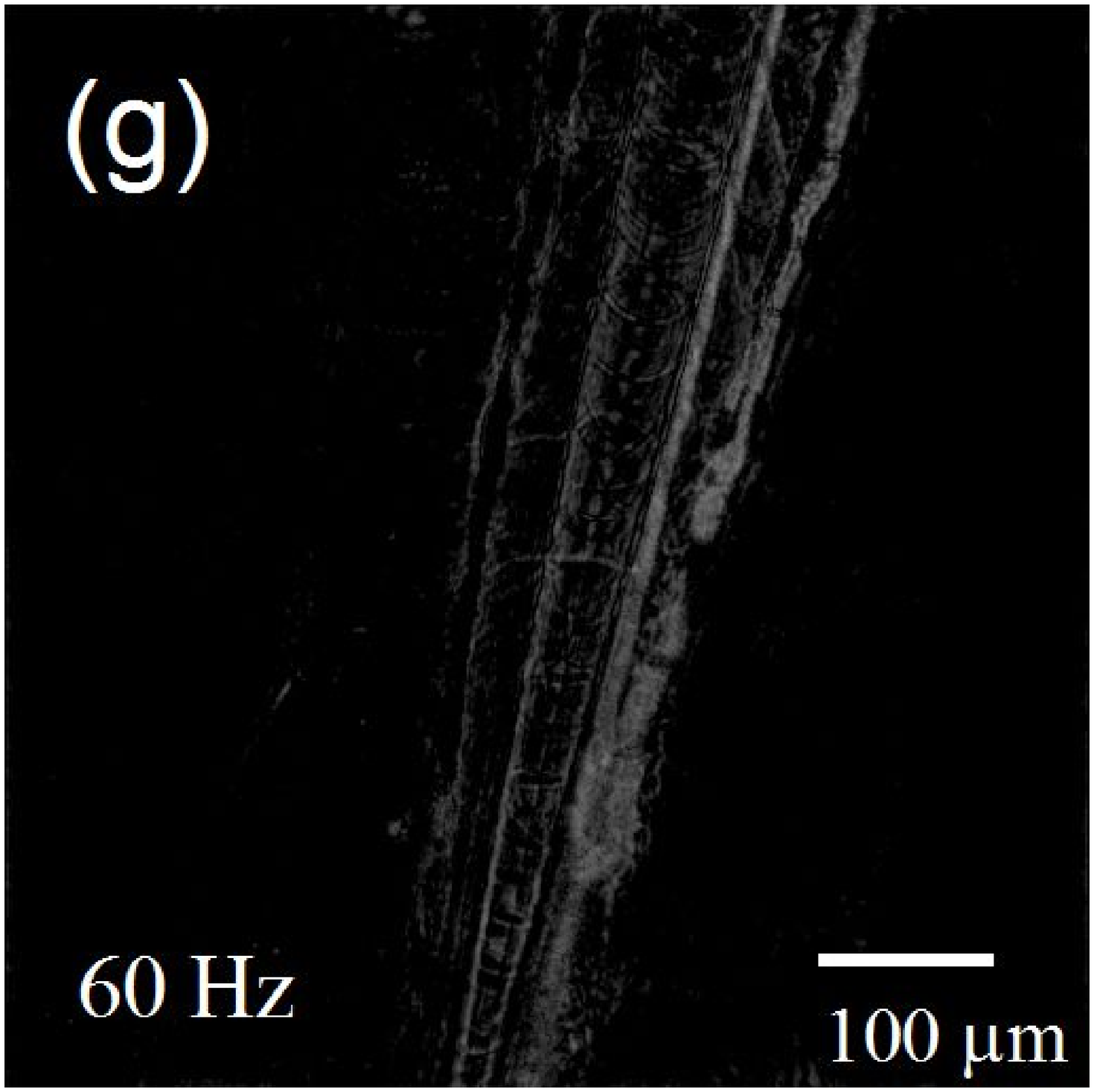}
\includegraphics[width = 3.4cm]{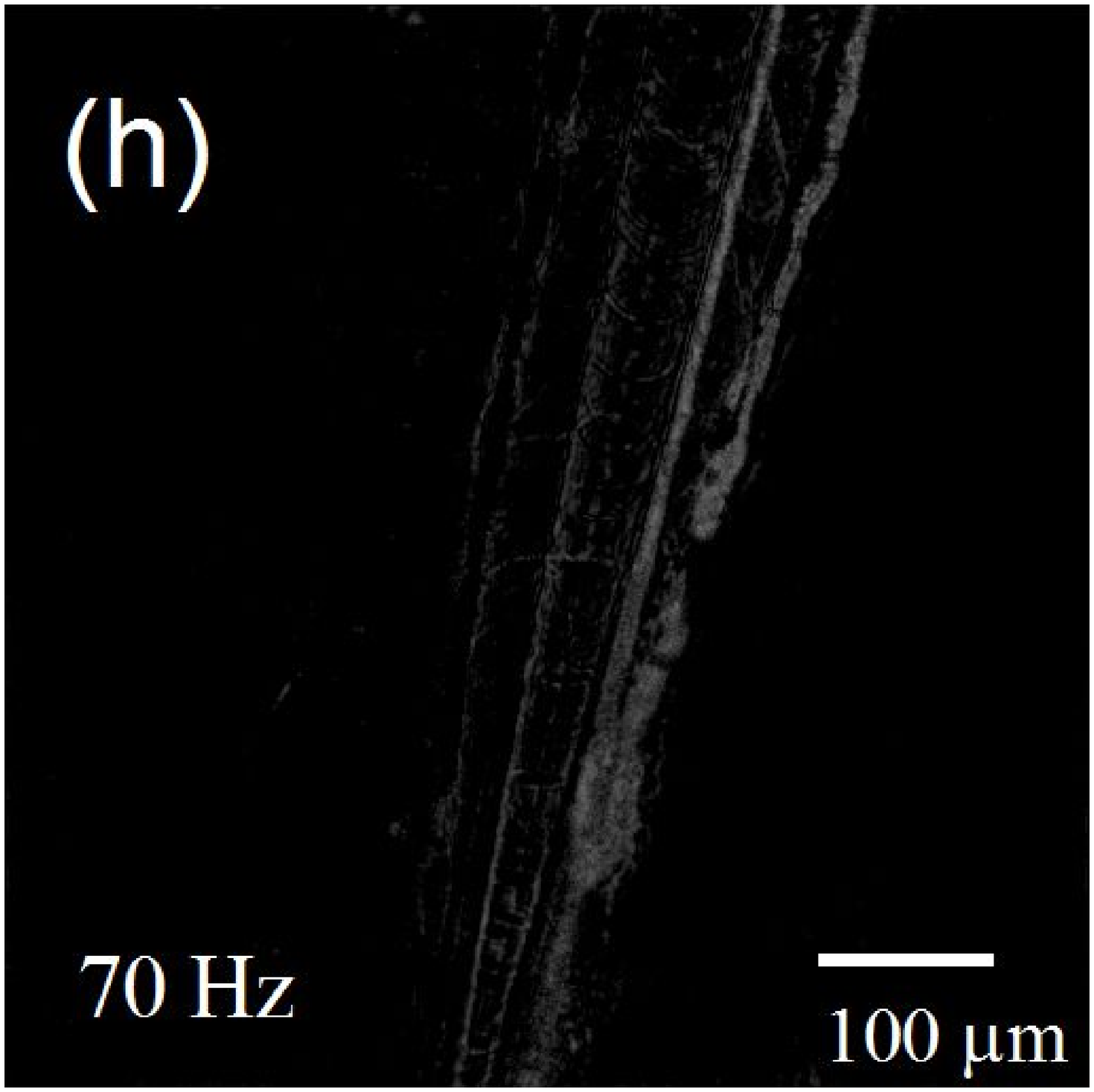}
\includegraphics[width = 3.4cm]{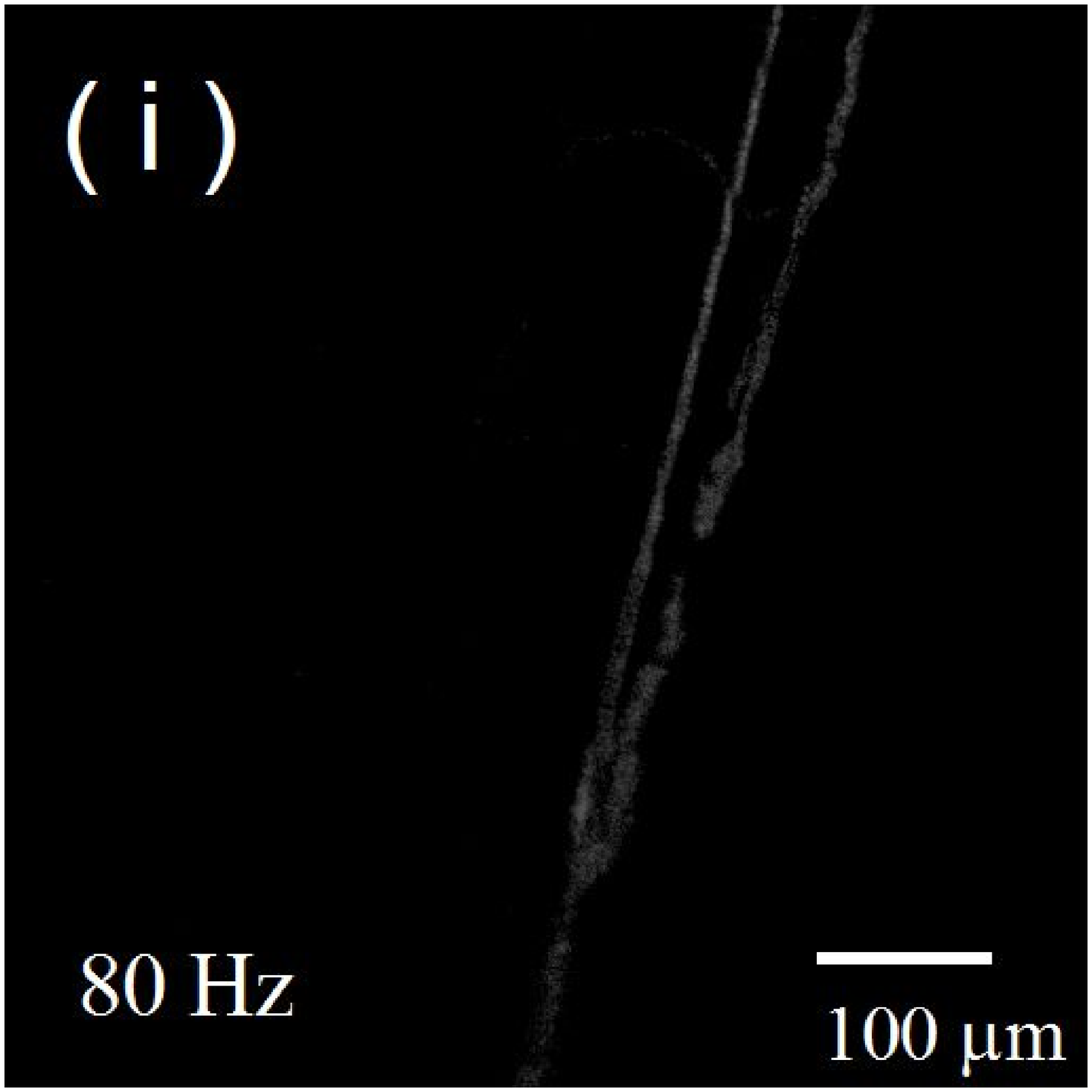}
\caption{Reconstructed hologram  $H_3(x,y,z=0)$ made for $(\omega_{LO}-\omega_I)/(2\pi)=
 0 $ Hz (a): see also \url{Media 2}, 10 Hz (b) ... to 80   Hz (i). Images (a) to (i) are displayed with the same Log scale for the average intensity  $\langle |H_{3}|^2\rangle$. The sample is the same as the one in Fig. \ref{fig_BW}.}
\label{fig_4}
\end{figure}

Figure \ref{fig_4} shows the Doppler  average reconstructed holograms $\langle |H_3(x,y)|^2\rangle $ obtained with frequency shifts $ \Delta \omega/(2\pi)$ varying from 0 Hz (a) to 80 Hz (i) by step of 10 Hz. The recorded hologram  $H(x,y)$, and the reconstructed intensity hologram $|H_3(x,y)|^2$ are calculated from successive sequence of 2 frames ($I_n$ and $I_{n+1})$  within the sequence of $N=32$ frames (with $N=0$ to  30), the  displayed signal $\langle |H_3(x,y)|^2\rangle $ corresponding    to the average of $|H_3(x,y)|^2$  over $n$. Since the ballistic peak is not seen, the contrast of the Fig. \ref{fig_4}  images is reversed with respect to Fig. \ref{fig_BW} (d), and vessels where blood flows are therefore seen in white on a black background. In image (a) caudal vein (A) and caudal artery (B,C) are seen as well as intersegmental vessels (D, E) and dorsal longitudinal  anastomotic vessels. An abnormal shunt between caudal artery and vein in seen near arrow (C).

\begin{figure}[htbp]
\centering
\includegraphics[height = 5cm]{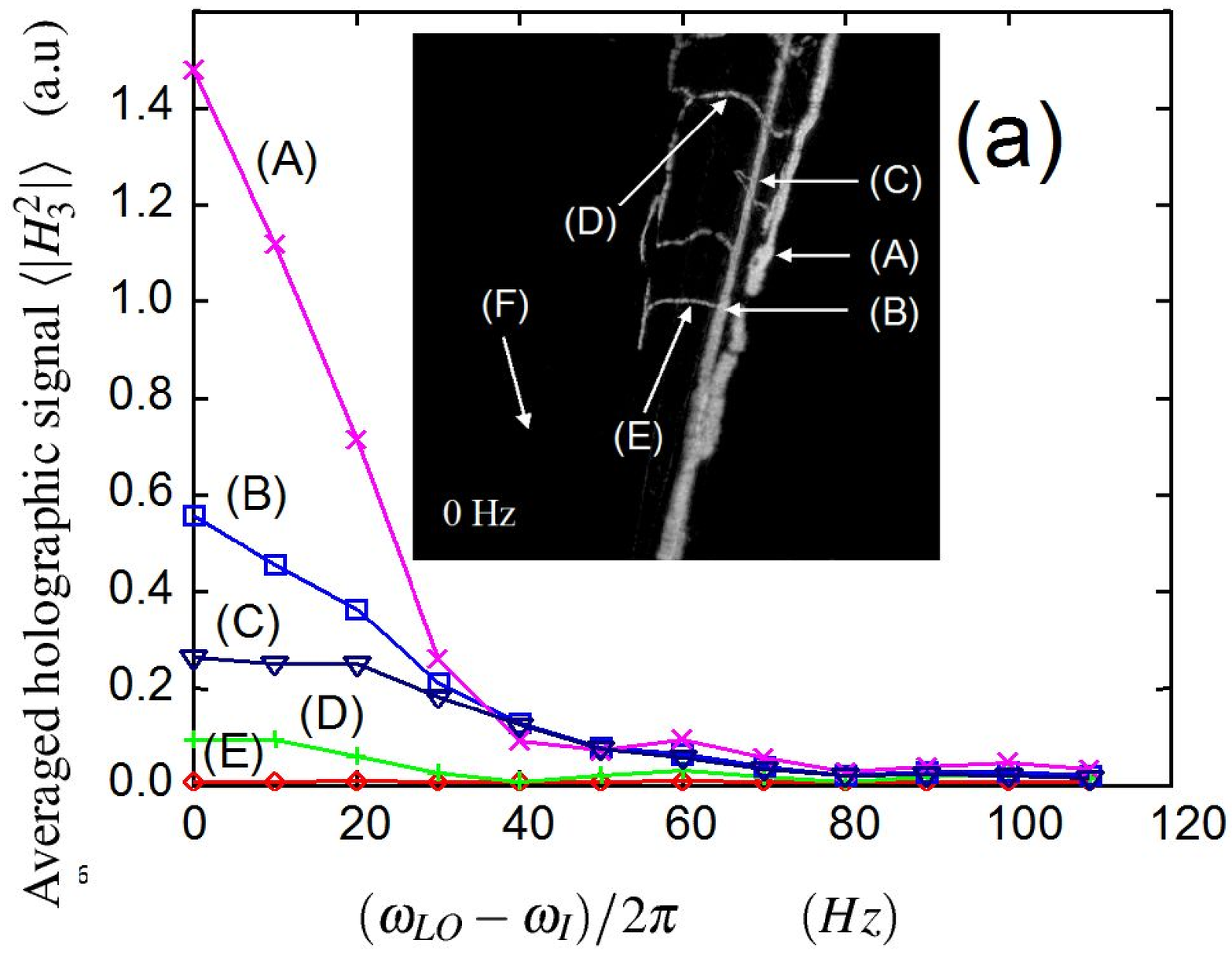}
\includegraphics[height = 5cm]{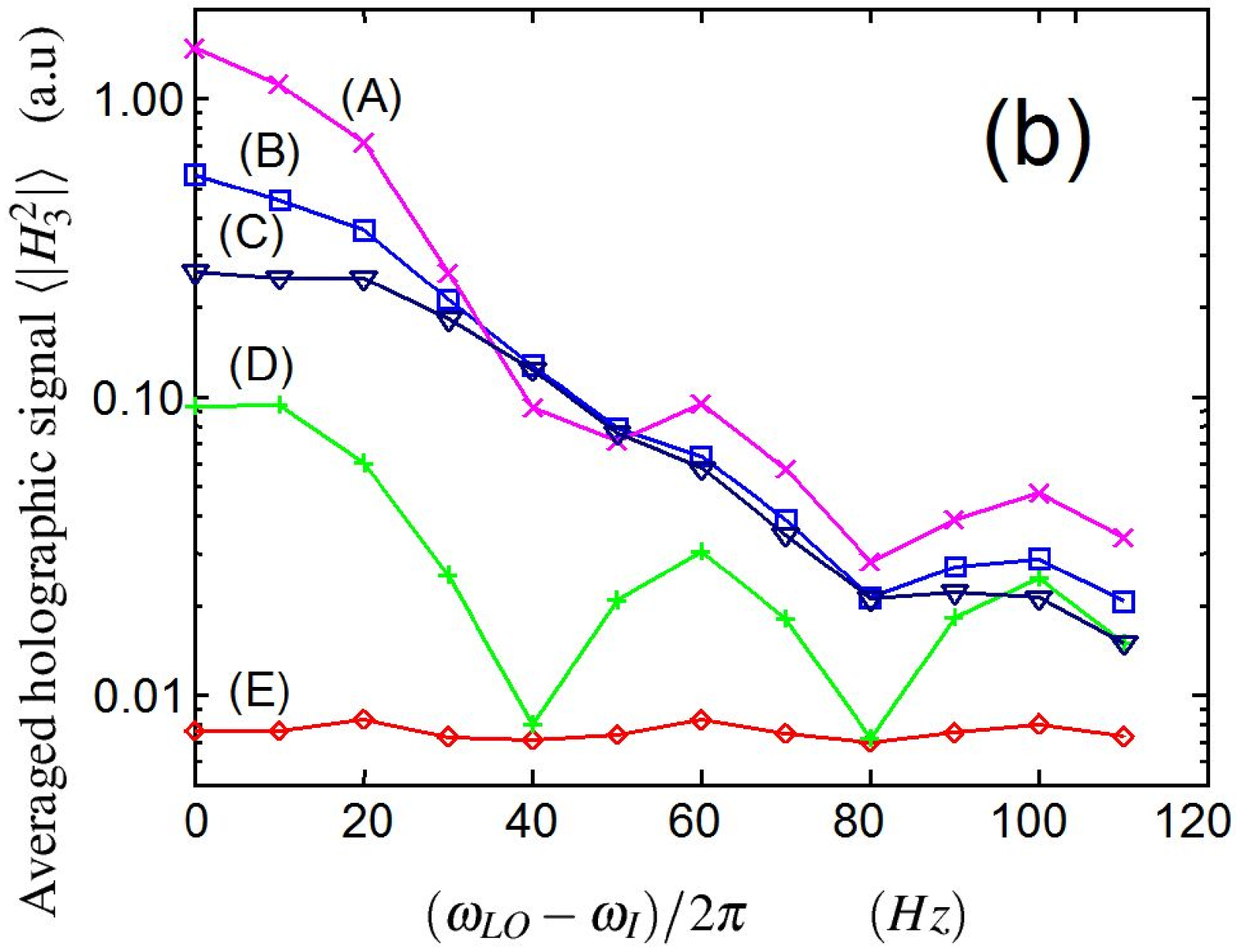}
\caption{Dependance of the Doppler holographic signal $\langle |H_3(x,y)|^2\rangle $ with the frequency offset $(\omega_{LO}-\omega_{\rm I})/(2\pi)$ for different location A to F  of the reconstructed image of Fig. \ref{fig_BW}(a). Curves are drawn with linear (a) and logarithmic (b) scales.}
\label{fig_curve}
\end{figure}

The reconstructed images (a) to (i) are displayed with the same Log scale for the average intensity  $\langle |H_{3}|^2\rangle$. One thus  clearly see the decrease of the holographic  signal  with the frequency shift  $(\omega_{LO}-\omega_{\rm I})/(2\pi)$ that is expected. Figure \ref{fig_curve} analyzes this decreases for the zones  of the reconstructed image marked by arrows (A) to (F) in Fig. \ref{fig_4}(a). All  curves are calculated by applying  on the holographic signal $\langle |H_{3}|^2\rangle$ a gaussian low pass filter (of 10 pixels width for curves A,B,C and F curves, and 4 pixels width for curves D and E), and by summing  the filtered holographic signal over $8\times 8$ pixels.
On curve (A), which corresponds to the flow in caudal vein, the holographic signal decreases regularly with the frequency offset $\Delta \omega$. The same occurs in (B) and (C),  but the decrease is slower showing faster velocities for the blood cells in the caudal artery.

%
%

In intersegmental vessels  (curves D and E), the signal  $\langle |H_{3}|^2\rangle$ is much lower than the one of larger artery and vein. The corresponding  frequency spectrum (Fig. \ref{fig_curve}D and E) exhibit  a decrease of the signal at $\Delta \omega/(2\pi)=40$ and 80 Hz offset followed by an increase at 60 and 100 Hz. This might be related to the instrumental response of our holographic detection \cite{AtlanGrossAbsil2007}.
A fine analysis shows that the signal fluctuates a lot within each sequence of $N$ frames, and from one sequence to the next, since the flowing blood cells  can be visualized individually on the instantaneous intensity signal $|H_{3}(t)|^2$  reconstructed with successive sequence of 2 frames. 
Movie \url{Media 2} shows $|H_{3}(t)|^2$ for the sequence of $N=32$ frames recorded at 0 Hz (i.e. without frequency offset) that has been  used to display $\langle |H_{3}|^2\rangle$ on Fig. \ref{fig_4}(a). Movie \url{Media 3} shows $|H_{3}(t)|^2$ for a sequence of $N=96$ frames that image the embryo shown in Fig. \ref{fig_5}(h) and (i). The RBCs individual motion in intersegmental vessels is clearly seen on \url{Media 2} and \url{Media 3}. These two movies are very similar to those seen in reference \cite{gao2012vivo}. They can be used to measure the RBC velocity by Particule Image Velocimetry (PIV).

\subsection{RGB holographic Doppler images yielding direction of  motion.}

\begin{figure}[htbp]
\centering
\includegraphics[width = 3.4cm]{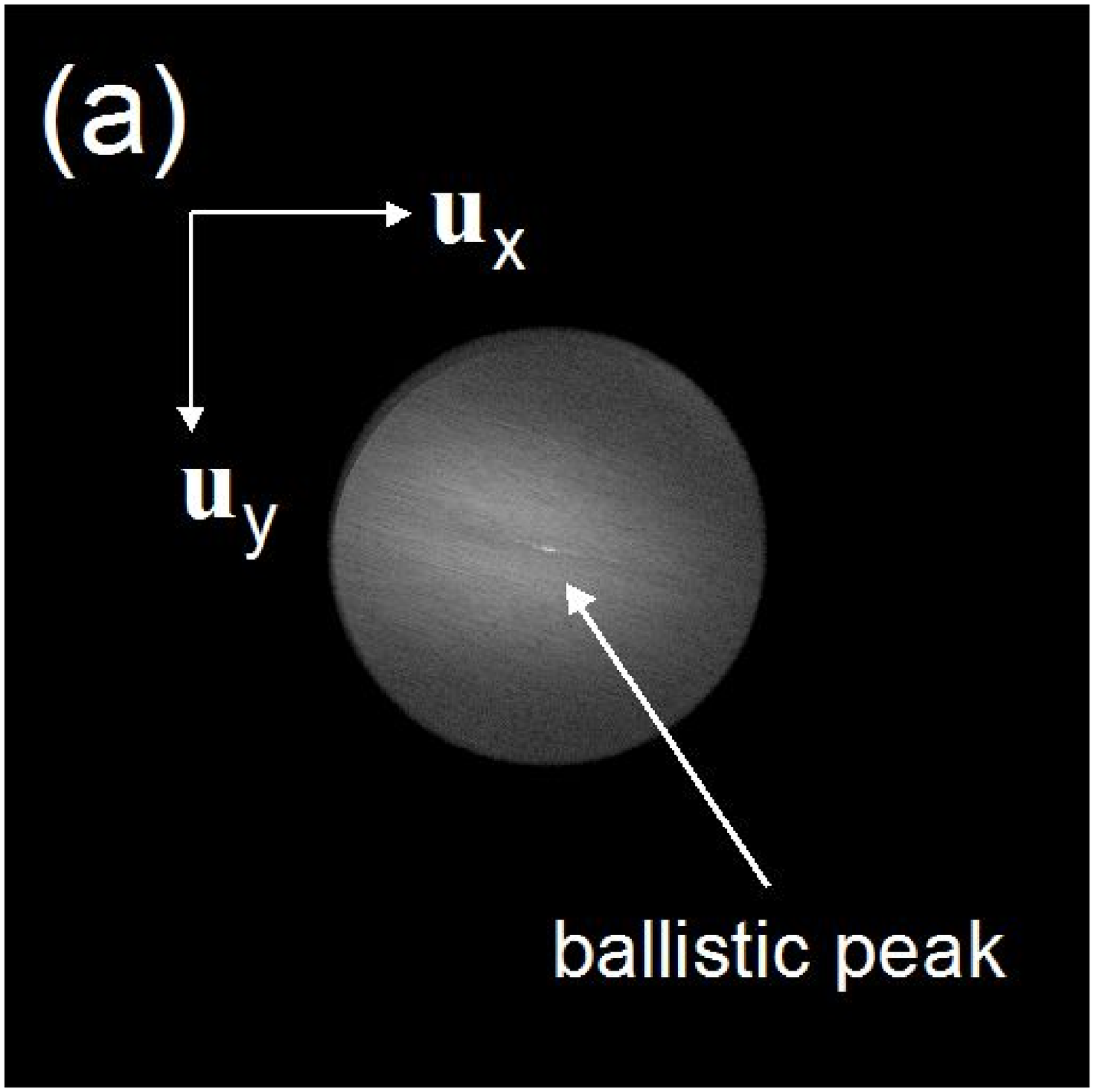}
\includegraphics[width = 3.4cm]{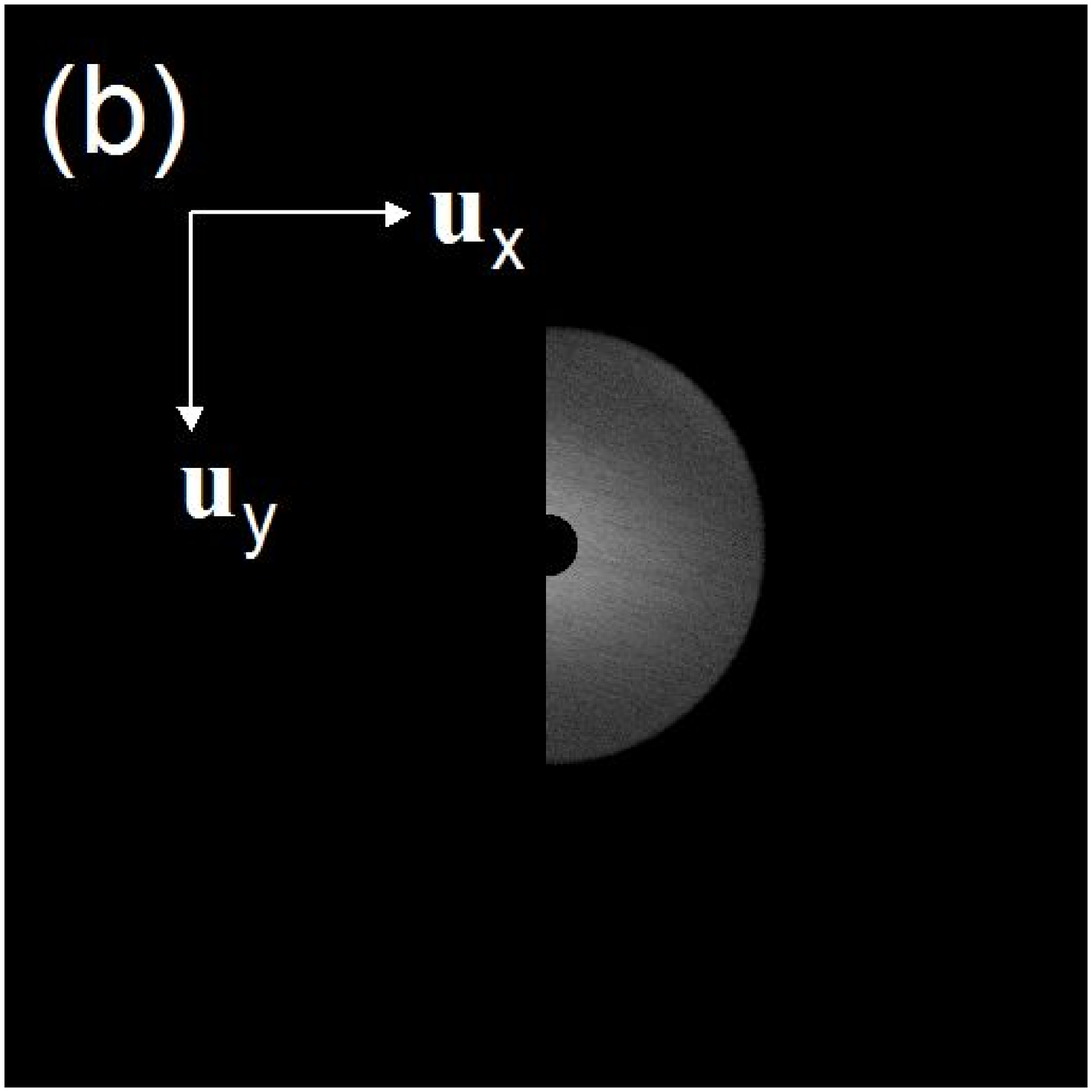}
\includegraphics[width = 3.4cm]{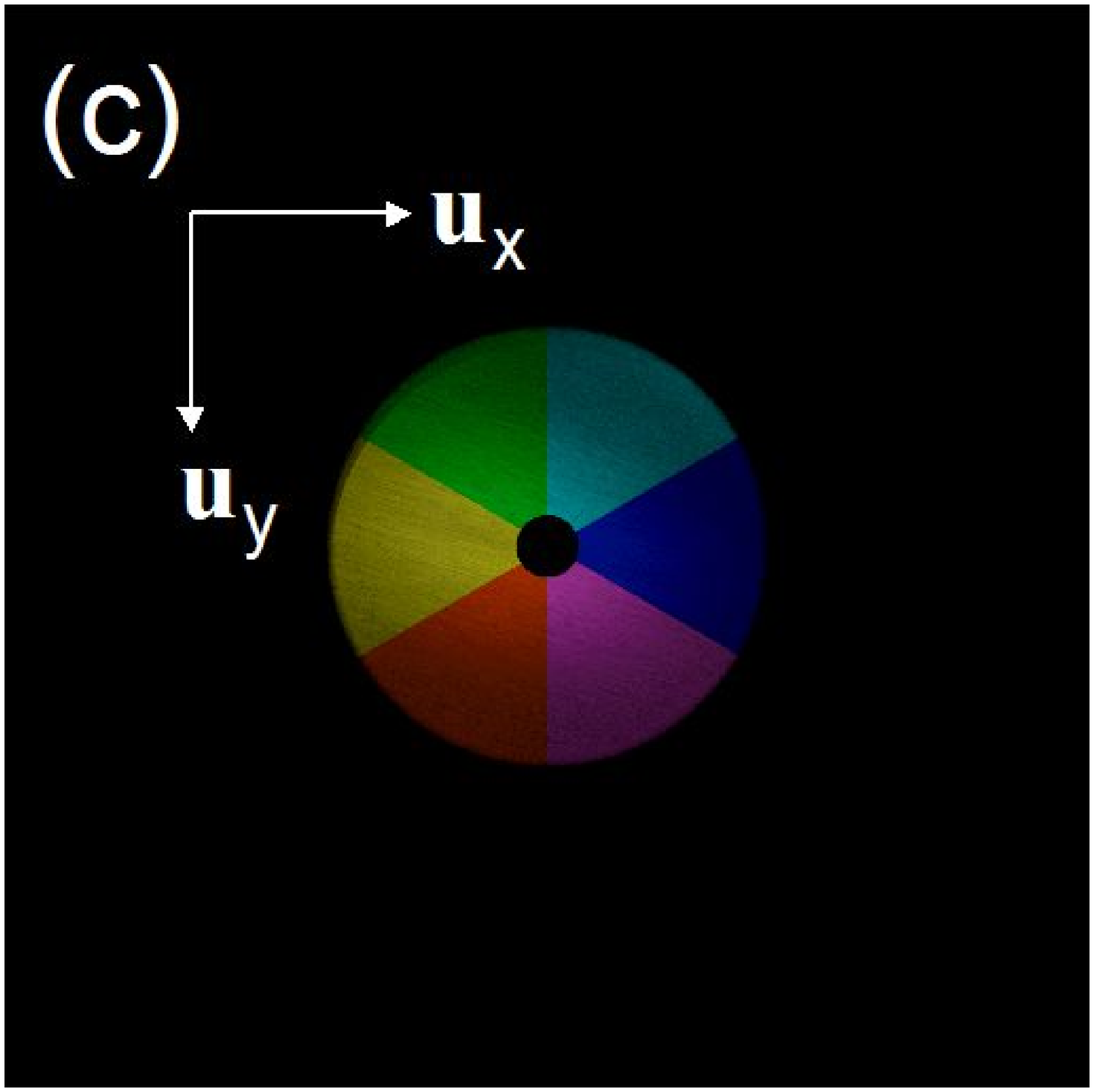}
\caption{Fourier space reconstructed hologram  $H_2(k_x,k_y)$ made without (a) and with (b,c) selection of the scattered wave vector $\textbf{k}_S$. In (b) the selected zone is oriented toward  $\textbf{u}_{\varphi=0}$. In (c), three zones oriented along  $\textbf{u}_{\varphi}$ with $\varphi=0$ (blue), $2\pi/3$ (green) and  $4\pi/3$ (red) are selected.  Display is made in arbitrary log scale for  $\langle |H_{2}|^2\rangle$. Frequency shift is $(\omega_{LO}-\omega_{\rm I})=0$ . }
\label{fig_5k}
\end{figure}

The holographic experiment made in transmission configuration gives here a Doppler signal corresponding to single scattering events like in Dynamic Light Scattering (DLS) experiments. The Doppler shift is thus simply  $\textbf{q}\textbf{.} \textbf{v}$ where $ \textbf{q} = \textbf{k}_S-\textbf{k}_{\rm I}$. The  illumination (or incident)  wave vector  $\textbf{k}_{\rm I}$ is fixed by the  geometry of the experimental setup. In  transmission geometry,   $\textbf{k}_{\rm I}$ is parallel to $ \textbf{ u}_z$ (where  $ \textbf{ u}_x $,  $ \textbf{ u}_y$ and $ \textbf{ u}_z$ are the unit vectors in the $x,y$ and $z$ directions). On the other hand, $\textbf{k}_S=(k_x, k_y,k_z)$ with $|\textbf{k}_S|=2\pi/\lambda $ is measured by our holographic experiment. In the reconstruction procedure, the cropped and translated  zone  seen on Fig. \ref{fig_BW}(c) and on Fig \ref{fig_5k}(a) is a map of the holographic signal as a function of $k_x$ and $k_y$.

It is then possible to  select   $\textbf{k}_S$ by spatial filtering   in the Fourier space, and to reconstruct  images corresponding to given intervals of $\textbf{k}_S$. To illustrate this idea, we have selected three $\textbf{k}_S$ zones covering  half of the Fourier space. These zones are   oriented toward $\textbf{u}_{\varphi=0}$, $\textbf{u}_{\varphi=2\pi/3}$ and $\textbf{u}_{\varphi=4\pi/3}$
with $ \textbf{u}_{\varphi}=\cos \varphi ~\textbf{u}_x + \sin \varphi ~\textbf{u}_y$. With the holographic signal $H_2(k_x,k_y)$ of each zone, we have reconstructed a Red, Green or Blue colored image $\langle|H_3(x,y,z)|^2 \rangle $, and we have combined these three images to get RGB images sensitive to the direction of $\textbf{k}_S$ (and thus to the direction of $\textbf{q}$ and $\textbf{v}$).  Figure \ref{fig_5k} (b) shows the    $|H_2(k_x,k_y)|^2$ signal, filtered by the $\varphi=0$ spatial filter, which corresponds to the   blue image. Figure \ref{fig_5k} (c) shows the Fourier space RGB image of the $|H_2(k_x,k_y)|^2$ signal obtained by combining the three colored  images corresponding to the $\varphi=0$, $\varphi=2\pi/3$ and $\varphi=4\pi/3$ filters.

One must notice here, that the wave vector of  ballistic light remains parallel to $\textbf{u}_z$ yielding a bright peak near the center of the Fourier space ($k_x,k_y\simeq 0$). Although most of the ballistic light is removed in 2 phase holograms recorded without frequency offset, this peak still remains   visible in that case as seen on  Fig.  \ref{fig_5k}(a). To fully eliminate  the ballistic light, the center of the Fourier space (i.e. $k_x,k_y\simeq 0$) has thus been  removed from the selected zones as shown on Figure \ref{fig_5k}(b) and (c).

\begin{figure}[htbp]
\centering
\includegraphics[width = 3.4cm]{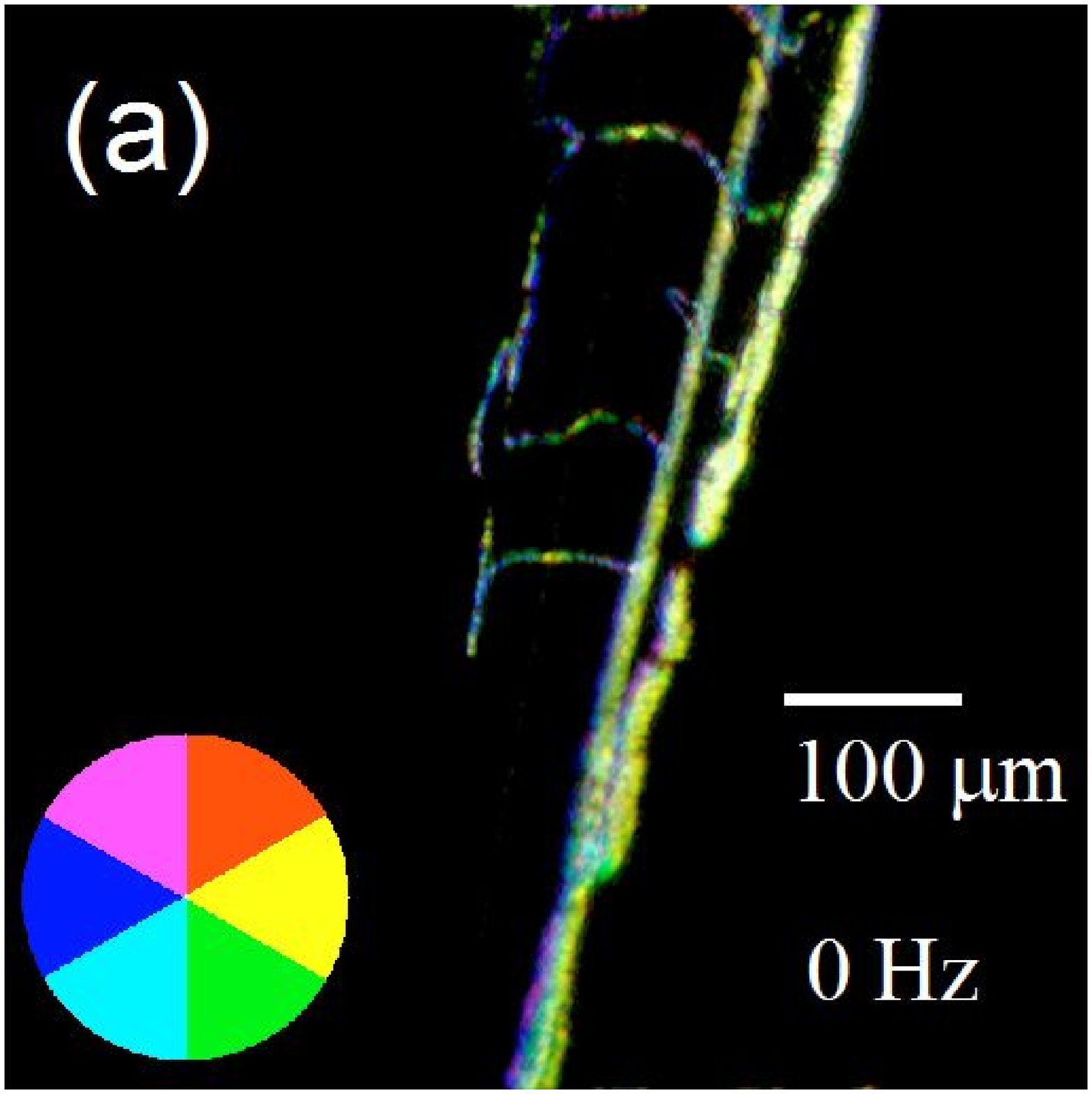}
\includegraphics[width = 3.4cm]{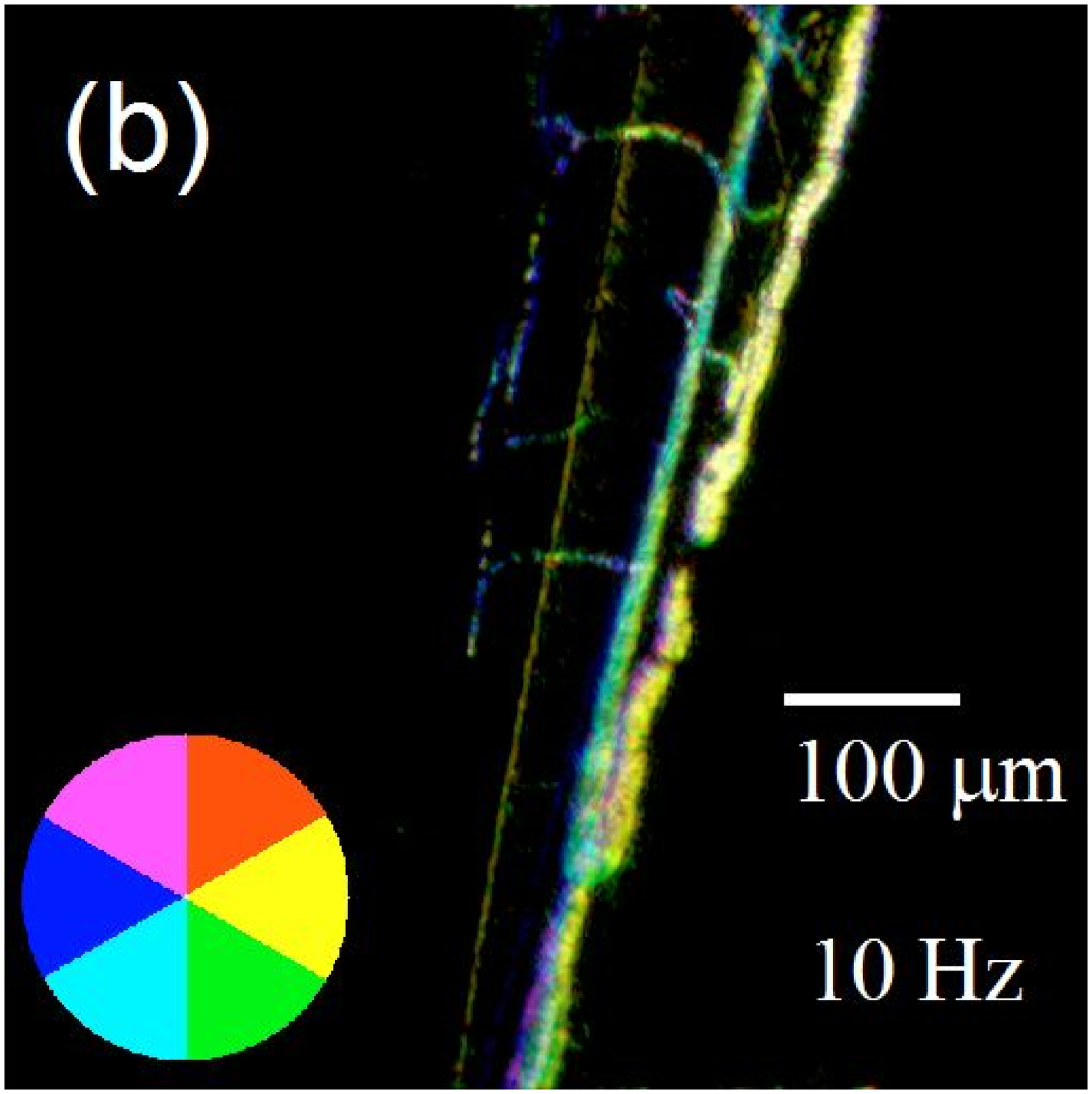}
\includegraphics[width = 3.4cm]{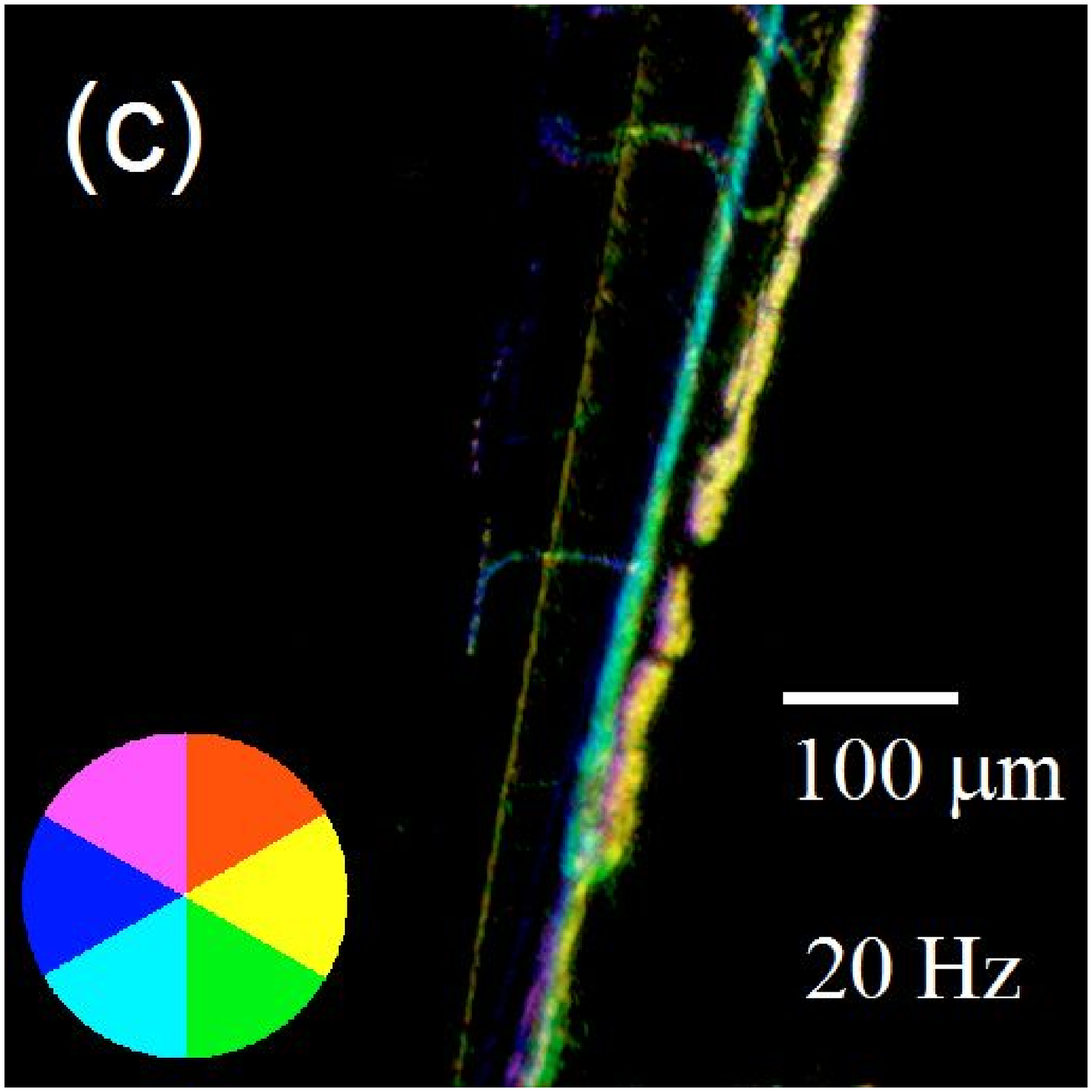}
\includegraphics[width = 3.4cm]{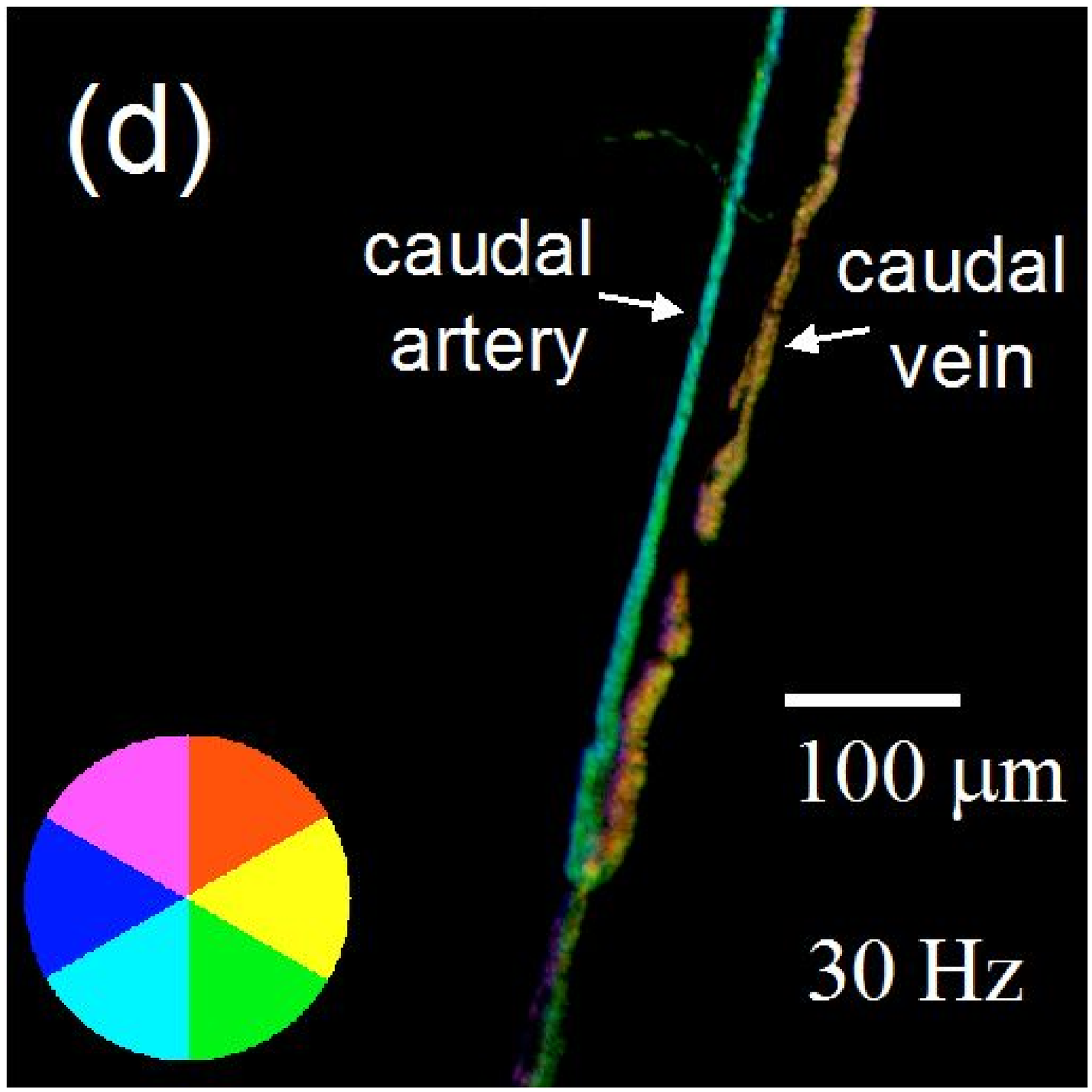}
\includegraphics[width = 3.4cm]{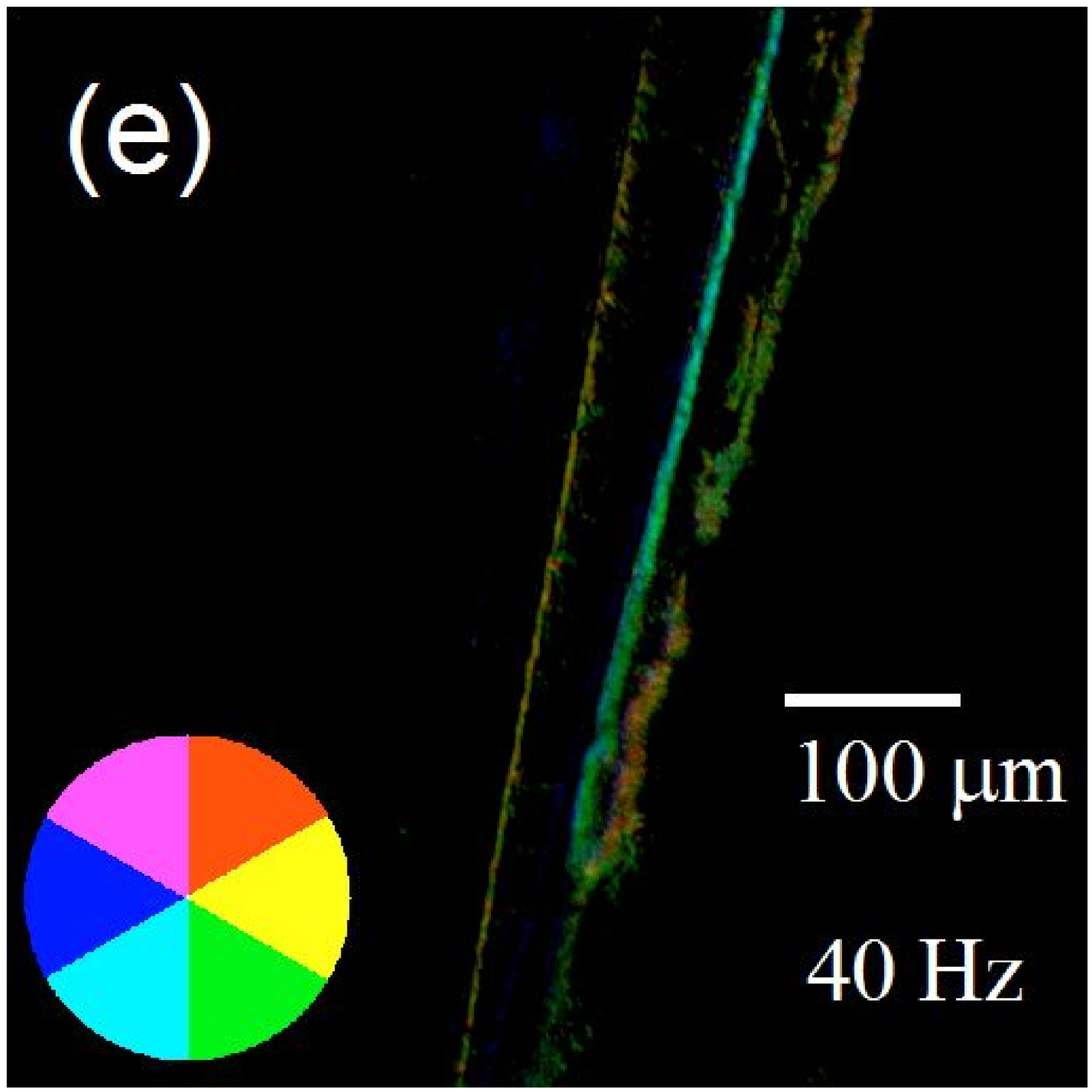}
\includegraphics[width = 3.4cm]{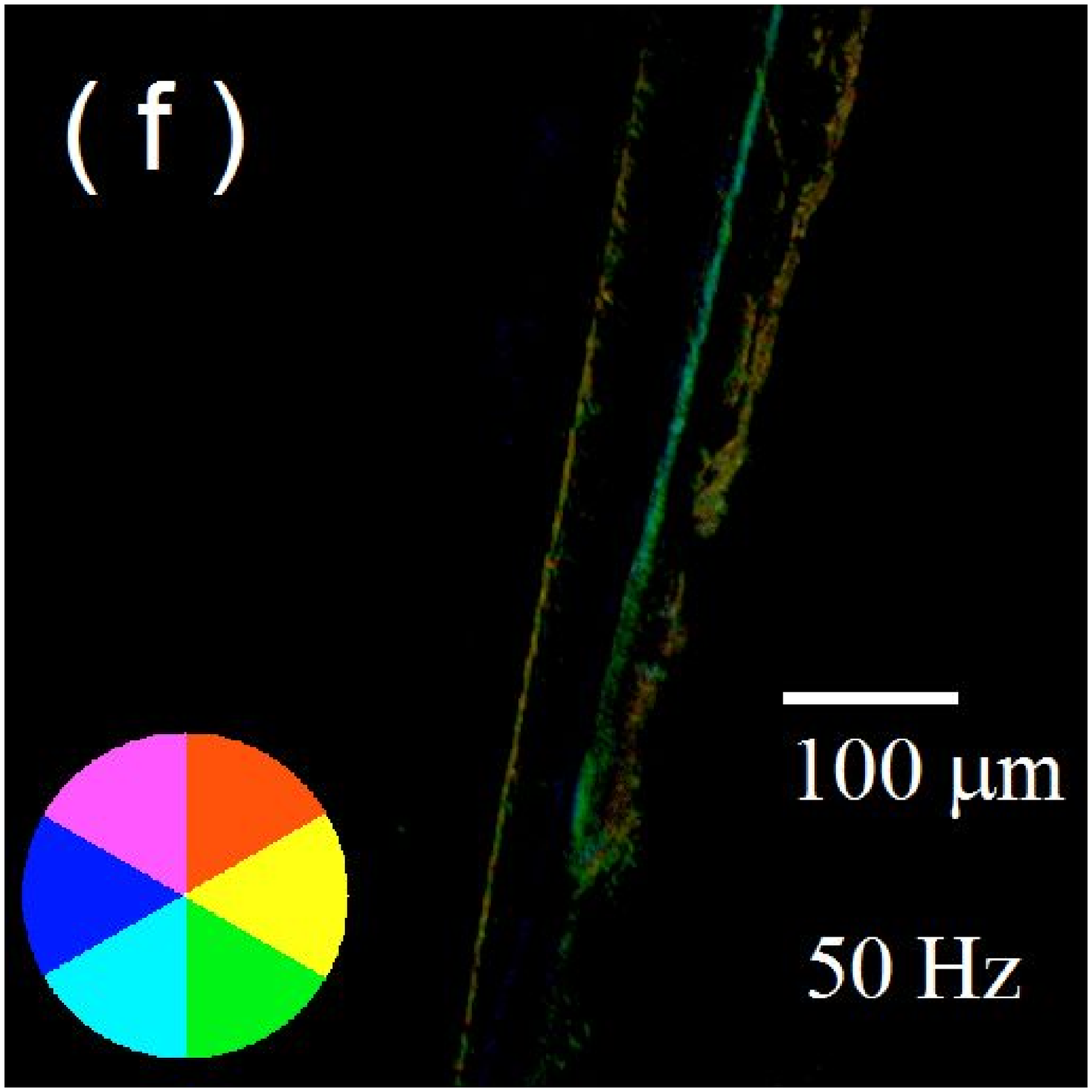}
\includegraphics[width = 3.4cm]{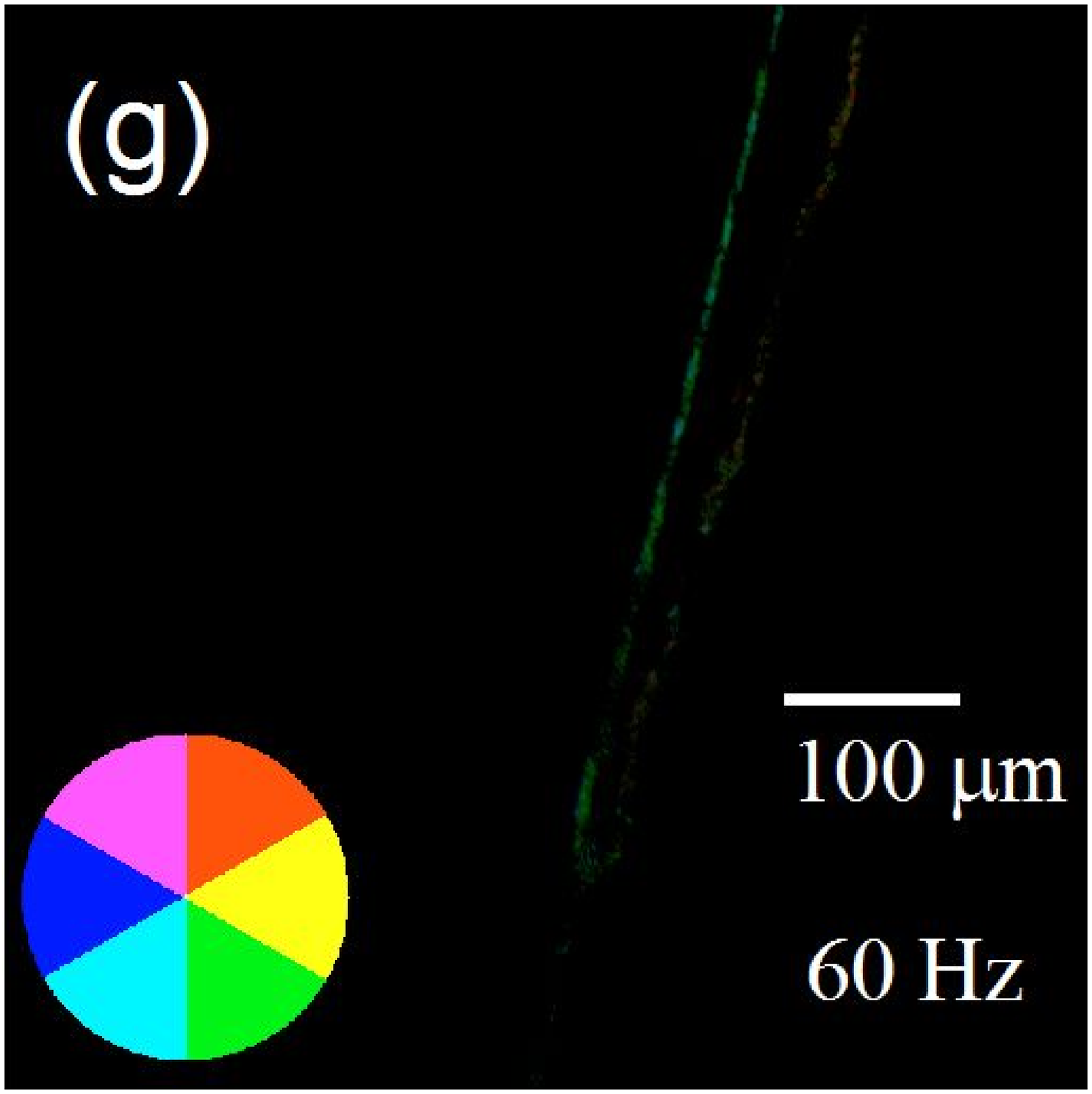}
\includegraphics[width = 3.4cm]{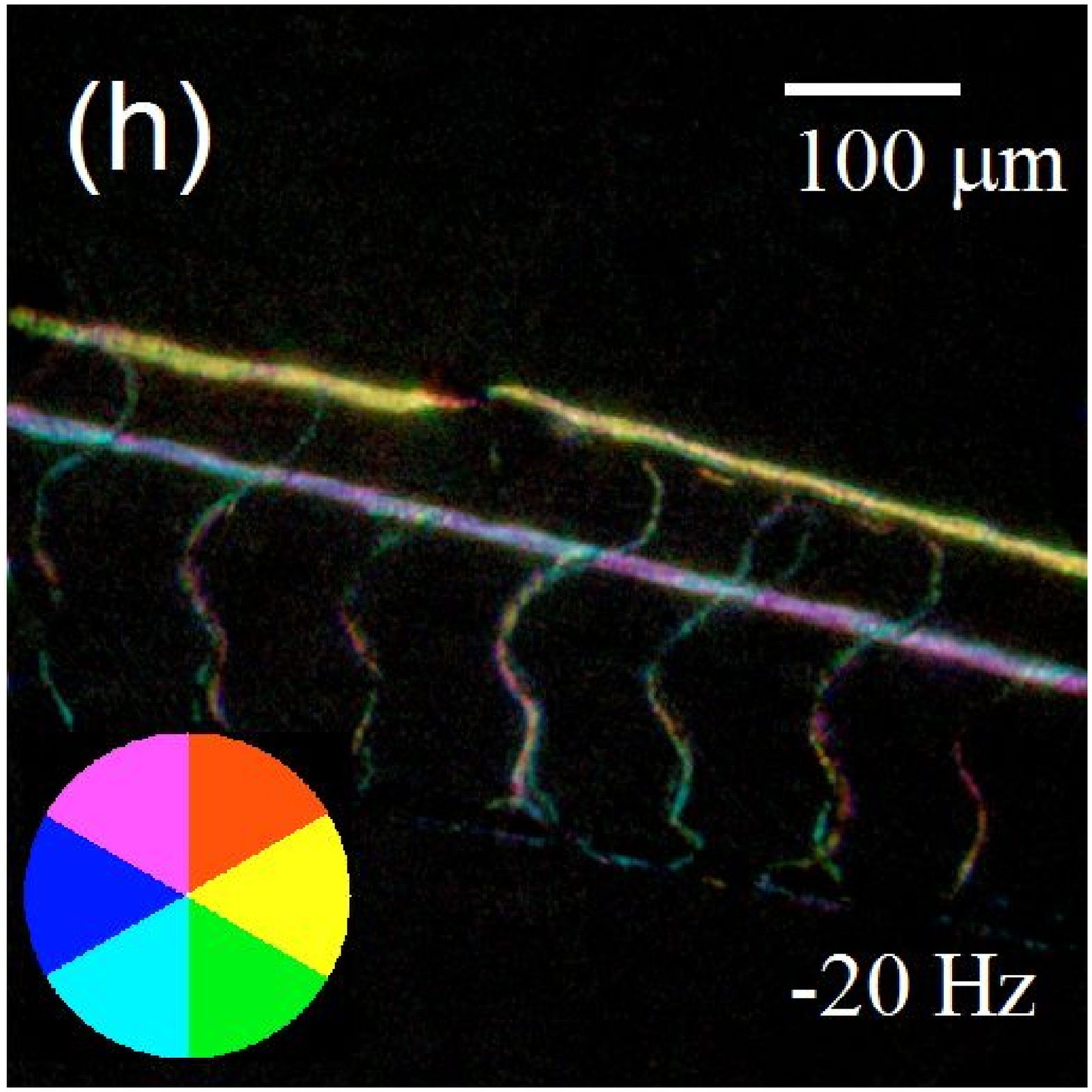}
\includegraphics[width = 3.4cm]{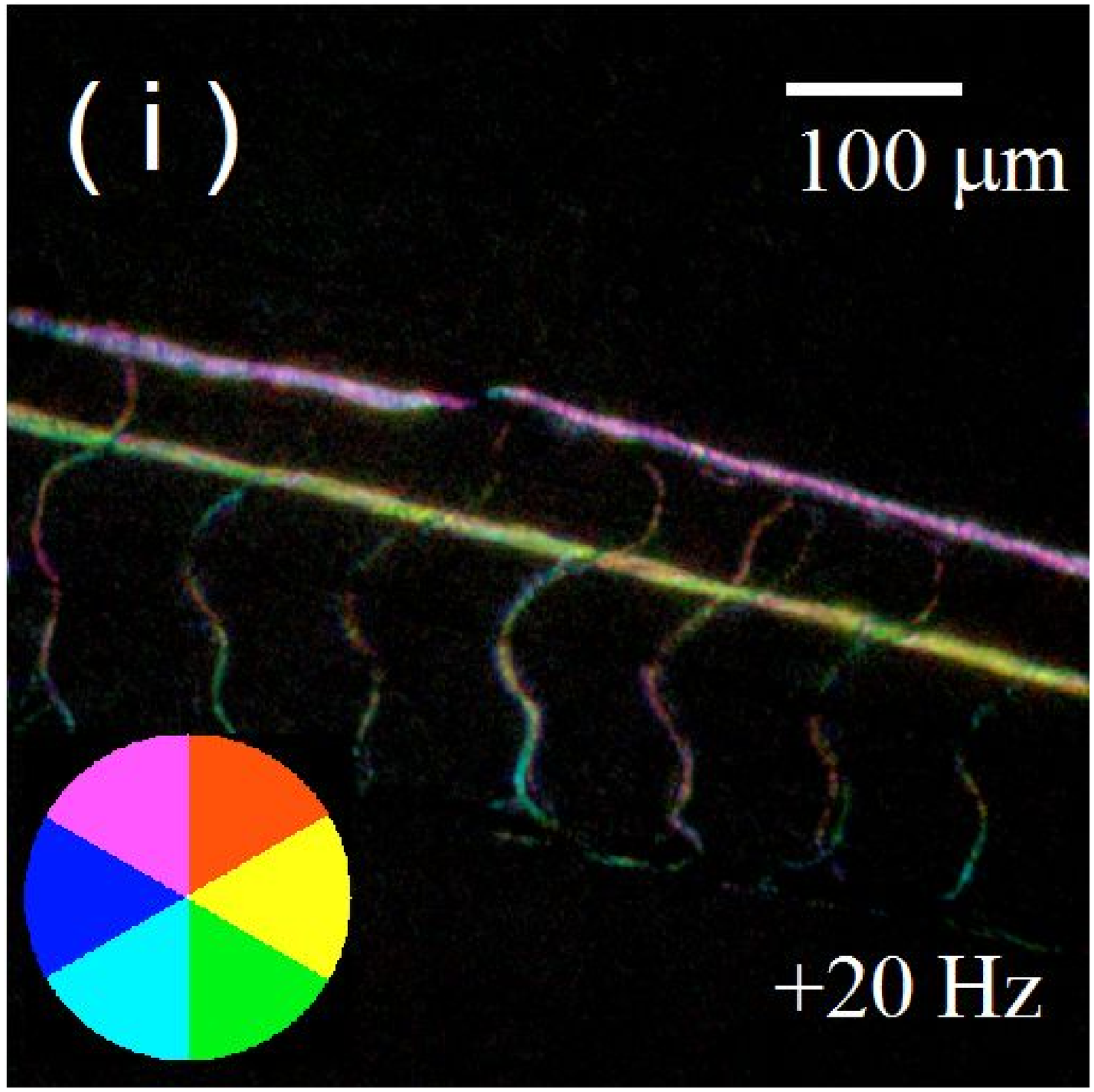}
\caption{Colored reconstructed hologram  $H_3(x,y,z=0)$ made for $(\omega_{LO}-\omega_I)/(2\pi)=
 0 $ Hz (a), 10 Hz (b) ... to 60   Hz (g), and for -20 Hz (h) and + 20 Hz (i). Images (a) to (g) are made with same zebrafish sample and  same viewpoint  as Fig. \ref{fig_BW} and Fig. \ref{fig_4}. Images (h) to (i) correspond to another fish embryo.
Images (a) to (g) (and (h) and (i)) are displayed with the same color RGB Log scale for the average intensity  $\langle |H_{3}|^2\rangle$. \url{Media 3} corresponds to images (h) and (i).}
\label{fig_5}
\end{figure}

Figure \ref{fig_5}(a) to (g) shows the RGB colored reconstructed image of the zebrafish sample obtained with the holographic data of Fig. \ref{fig_4}(a) to (g). Without frequency offset (Fig.  \ref{fig_5}(a)  with $\Delta \omega=0$), the RGB colored  reconstructed image is very similar to the black and white one (Fig. \ref{fig_4}(a)), but with lower resolution, since  half of the Fourier space information is used in the reconstruction. The images of Fig. \ref{fig_5} are obtained from a sequence of 32 images (i.e. with an acquisition time of 3.2 s). The reconstruction  is made on GPU in less than one second for the 32 images.
Since the detection shift $\Delta \omega$  is null, the detection efficiency does not depend on the sign of the Doppler shift of the signal $\omega_S-\omega_{\rm I}$. Detection is not sensitive to the sign of $\textbf{q}\textbf{.} \textbf{v}$, nor to the $\textbf{v}$ direction (even if  $\textbf{q}$ direction is selected in the Fourier space). The colors in the reconstructed image of Fig.  \ref{fig_5}(a) are thus the same whatever the motion direction is. With increasing frequency shifts $\Delta \omega$, signal decreases but colors change, depending on the flow direction. On figure \ref{fig_5}(d), with $\Delta \omega = 30 $ Hz, caudal artery appears in cyan, while caudal vein, where blood flows in opposite direction, is seen  in orange. The flow direction corresponds  here to the Fig. \ref{fig_5} color wheel. This wheel corresponds to the colored image of the Fourier space filtered signal $|H_2(k_x,k_y)|^2$ of Fig. \ref{fig_5k}(d) rotated by 180$^\circ$.

To verify that the RGB colored images give a realistic information on the direction of the velocity $\textbf{v}$, we have recorded, with another embryo, holograms for both sign of the frequency offset $\Delta \omega$.   Figure \ref{fig_5}(h) and (i) shows the  reconstructed   RGB images obtained for $\Delta \omega /(2\pi)= \pm 20$ Hz. Here again, the colors of the reconstructed image depend on the flow direction, but the color coding is reversed for negative frequency offsets: $\Delta \omega <0$. The correspondence of the motion direction with the color wheel  is only valid for positive shifts $\Delta \omega >0$.

\section{Conclusion}

In this paper, we have shown that an upright commercial bio-microscope can be transformed into a powerful holographic setup,  by coupling the microscope  to an heterodyne holographic \cite{LeClerc2000} breadboard with optical fibers. By  controlling  the frequency offset $\Delta \omega$ of the reference beam,  and by using different combinations   of frames to calculate the hologram $H$, we have imaged a living zebrafish  embryo under different modalities. With $\Delta \omega=0$ and $H=(I_0-I_2)+j(I_1-I_3)$, we have calculated the amplitude  and the phase of the transmitted light. Although the phase can be qualitatively explored, we have not displayed phase images because the phase variation is too fast (the thickness of the embryo is  hundreds of microns).

With $\Delta \omega = m \omega_{CCD}$ (where  $m \ne 0$ is integer) and $H=(I_0-I_1)$, we have selected the signal whose Doppler shift  $\omega_S-\omega_{\rm I}$ is close to $ \Delta \omega$ according to Eq. \ref{fig_eta2} and Fig. \ref{fig_eta2}) and imaged the blood flow.  With $\Delta \omega = 0$ and $H=(I_0-I_1)$, we have visualized  individual moving bloods cells in intersegmental vessels, getting movies that show their instantaneous motion.

Since the  scattering is moderate like in DLS (Dynamic Light Scattering) experiment, the Doppler shift is directly related to the  scatterers (i.e. blood cells) velocity $\textbf{v}$ by  $\omega_S-\omega_{\rm I}=\textbf{q} \textbf{.} \textbf{v}$ with $\textbf{q}=\textbf{k}_S-\textbf{k}_{\rm I}$. Since $\textbf{k}_{\rm I}\parallel \textbf{u}_z$ is known, and since $\textbf{k}_S$ can be selected in the Fourier space, the scatterers velocity $\textbf{v}$ can be measured quantitatively by a proper choice of both  the frequency offset $\Delta \omega$ and   the Fourier space selected zone. To illustrate one of the numerous possibilities of Doppler holography in transmission, we have imaged  the blood cell motion direction  in RGB colors.

In the present paper, we have not used the possibilities for numerical refocusing to image different $z$ layers of our sample. This could be useful,  since all blood vessels are not in the same $z$ plane.  For instance, in Fig. \ref{fig_4} and Fig. \ref{fig_5}(a)-\ref{fig_5}(g), the caudal artery is roughly on focus, while the caudal vein is out of focus by about 100 $\mu$m. Refocusing could thus be used to reconstruct the embryo vascular system in 3D.

We  intend to further improve the method  by using a faster camera (to better analyze the motion of individual blood cells)  and a higher numerical aperture objective (for better optical sectioning).

\section*{Acknowledgments}
We  acknowledge   OSEO-ISI Datadiag grant and  ANR Blanc Simi 10 (n$^\circ$ 11 BS10 015 02) grant for funding, and R\'{e}my Vialla for technical support.

\end{document}